\documentclass[journal,twoside,web]{ieeecolor}
\usepackage{tmi}
\usepackage{cite}
\usepackage{amsmath,amssymb,amsfonts,multirow,tabularx,booktabs}

\DeclareMathOperator*{\argmin}{arg\,min}
\usepackage{algorithmic}
\usepackage{graphicx}
\usepackage{textcomp}
\definecolor{revision}{rgb}{0.,0.,0.}
\usepackage{hyperref}
\def\BibTeX{{\rm B\kern-.05em{\sc i\kern-.025em b}\kern-.08em
    T\kern-.1667em\lower.7ex\hbox{E}\kern-.125emX}}
\begin{document}
\title{Learning Whole Heart Mesh Generation From Patient Images For Computational Simulations}
\author{Fanwei Kong, Shawn C. Shadden
\thanks{This work was supported by the NSF, Award No. 1663747. We thank Drs. Shone Almeida, Amirhossein Arzani and Kashif Shaikh for providing the time-series CT image data. }
\thanks{Fanwei Kong, is with the Department of Mechanical Engineering, University of California at Berkeley,
Berkeley, CA 94720 USA (e-mail: fanwei\_kong@berkeley.edu).}
\thanks{Shawn C. Shadden, is with the Department of Mechanical Engineering, University of California at Berkeley,
Berkeley, CA 94720 USA (e-mail: shadden@berkeley.edu).}}

\maketitle

\begin{abstract}
Patient-specific cardiac modeling combines geometries of the heart derived from medical images and biophysical simulations to predict various aspects of cardiac function. However, generating simulation-suitable models of the heart from patient image data often requires complicated procedures and significant human effort. We present a fast and automated deep-learning method to construct simulation-suitable models of the heart from medical images. The approach constructs meshes from 3D patient images by learning to deform a small set of deformation handles on a whole heart template. For both 3D CT and MR data, this method achieves promising accuracy for whole heart reconstruction, consistently outperforming prior methods in constructing simulation-suitable meshes of the heart. When evaluated on time-series CT data, this method produced more anatomically and temporally consistent geometries than prior methods, and was able to produce geometries that better satisfy modeling requirements for cardiac flow simulations. Our source code and pretrained networks are available at \url{https://github.com/fkong7/HeartDeformNets}.
\end{abstract}


\section{Introduction}
\label{sec:introduction}

\IEEEPARstart{I}{mage}-based cardiac modeling is used to simulate various aspects of cardiac function, including electrophysiology \cite{Trayanova2011}, hemodynamics \cite{Mittal} and tissue mechanics \cite{Marx2020}. This method derives geometries of the heart from patient image data and numerically solves mathematical equations that describe various physiology on discretized computational domains. Such ``digital twin'' modeling of a patient's heart can provide information that cannot be readily measured to facilitate diagnosis and treatment planning \cite{Karabelas2018, Prakosa2018, KUNG2013423}, or to quantify biomechanical underpinnings of diseases \cite{MCDOWELL2012640_AF}. This paradigm has motivated numerous research efforts on a wide range of clinical applications, such as, simulations of the stress and strain of cardiac tissues when interacting with implantable cardiac devices\cite{MITRACLIP}, the cardiac flow pattern after surgical corrections \cite{Karabelas2018,KUNG2013423}, and cardiac rhythm outcome after ablation surgery \cite{Prakosa2018}. 

Generating simulation-suitable models of the heart from image data has remained a time-consuming and labor-intensive process. It is the major bottleneck limiting large-cohort validations and clinical translations of functional computational heart modeling\cite{Mittal, Strocchi2020}. Indeed, prior studies have been limited to only a few subjects\cite{Mittal,Karabelas2018,Prakosa2018}. The entwined nature of the heart makes it difficult to differentiate individual cardiac structures, and typically  a complicated series of steps are needed to identify and label various structures for the assignment of boundary conditions or modeling parameters. Deforming-domain computational fluid dynamics (CFD) simulations of the intracardiac hemodynamics, is particularly labor-intensive since it requires reconstructing temporally-consistent deformations of the heart from a sequence of image snapshots.  

Deep learning methods can train neural networks from existing data to automatically process medical images and generate whole heart reconstructions. Most deep learning methods have, however, focused on segmentation (pixel classification) rather than construction of a computer model of the heart, usually represented by tessellated meshes\cite{ZHUANG2019}. Prior studies on automated cardiac mesh reconstruction thus adopted multi-stage approaches, where segmentation of the heart was first generated by convolutional neural networks (CNN) and surface meshes of the heart were then created from the marching cube algorithms and following surface post processing methods\cite{KONG2020}. However, the intermediate segmentation steps often introduce extraneous regions containing topological anomalies that are unphysical and unintelligible for simulation-based analyses \cite{KONG2020}. Direct mesh reconstruction using geometric deep learning \cite{Defferrard2016, Bronstein2017} provides a recent avenue to address the end-to-end learning between volumetric medical images and simulation-ready surface meshes of the heart \cite{Voxel2Mesh, kong2021deeplearning, Attar2019}. However, these approaches often assumes the connectivity of the meshes. That is, the shape and topology of the predicted meshes from these approaches are pre-determined by the mesh template and cannot be easily changed to accommodate various mesh requirements for different cardiac simulations.

To overcome these short comings, we propose to learn to deform the space enclosing a whole heart template mesh to automatically and directly generate meshes that are suitable for computational simulations of cardiac function. Here we propose to leverage a control-handle-based shape deformation method to parameterize smooth deformation of the template with the displacements of a small set of control handles and their biharmonic coordinates. Our approach learns to predict the control handle displacements to fit the whole heart template to the target image data. We also introduce learning biases to produce meshes that better satisfy the modeling requirements for computational simulation of cardiac flow. The contributions of this work are summarized as follows:

\begin{enumerate}
    \item We propose a novel end-to-end learning method combining deformation handles to predict deformation of whole heart mesh templates from volumetric patient image data. We show that our approach achieves comparable geometric accuracy for whole heart segmentation as prior state-of-the-art segmentation methods.
    \item We introduced novel mesh regularization losses on vessel inlet and outlet structures to better satisfy the meshing requirements for CFD simulations. Namely, our method predicts meshes with coplanar vessel caps that are orthogonal to vessel walls for CFD simulations. 
    \item We validated our method on creating 4D dynamic whole heart and left ventricle meshes for CFD simulation of cardiac flow. Our method can efficiently generate simulation-ready meshes with minimal post-processing to facilitate large-cohort computational simulations of cardiac function.
\end{enumerate}
\subsection{Learning-Based Shape Deformation}
Shape deformation using low-dimensional control of deformation fields has been extensively studied for decades in computer graphics and has been ubiquitously used in animation. These methods usually interpolate the transformation of a sparse set of control points to all points on the shape. Among the most popular approaches were are free-form-deformation that uses a regular control point lattice to deform the shape enclosed within the lattice \cite{Sederberg1986FreeformDO}, cage-deformation that uses a convex control cage that encloses the shape \cite{Nieto2013CageBD}, as well as control-handle-based approaches that directly place control points on the surface of the shape \cite{SorkineHornung2004LaplacianSE, Jacobson2014BoundedBW, Wang2015LinearSD}. Recent works have shown success in integrating these shape deformation methods in deep-learning frameworks for automated mesh reconstruction from single-view camera images\cite{Kurenkov2018DeformNetFD}, generative shape modeling\cite{Liu2021DeepMetaHandlesLD} as well as deformation transfer \cite{Wang2020NeuralCF}. However, these approaches were designed to take 2D camera images or 3D meshes as input and used memory-intensive CNNs or fully connected neural network to predict the transformation of control points. They thus cannot be directly applied to deform complicated whole heart structures from high-resolution 3D medical image data. Therefore, we herein propose to use graph convolutional networks (GCN) and sparsely sampling of the volumetric image feature map to predict control point translations and thus efficiently produce meshes from 3D medical images.  

\subsection{Mesh Reconstruction From 3D Medical Images}
Recent works on direct mesh reconstruction from volumetric images aim to deform an initial mesh with pre-defined topology to a target mesh \cite{kong2021deeplearning, Voxel2Mesh}. Our previous approach leveraged a GCN to predict deformation on mesh vertices from a pre-defined mesh template to fit multiple anatomical structures in a 3D image \cite{kong2021deeplearning}. However, different structures were represented by decoupled mesh templates and thus still required post-processing to merge different structures for computational simulations involving multiple cardiac structures. Similarly, \cite{Attar2019} used deep neural networks and patient metadata to predict cardiac shape parameters of a pre-built statistical shape model of the heart. Our approach presented herein, in contrast, deforms the space enclosing the mesh template. Once being trained on the whole heart template, our network can deform alternative template meshes that represent a subset of the geometries in the template to accommodate different modeling requirements. 

A few studies have focused on learning space deformation fields. \cite{Pak2021DistortionEF} used a 3D UNet to predict a deformation field to deform heart valve templates from CT images. Additionally, our preliminary work combined free-form deformation (FFD) with deep learning to predict the displacement of a control point grid to deform the space enclosing a simulation-ready whole heart template\cite{KONG2021FFD}. However, predicting the deformation fields requires many degrees of freedom to produce accurate results. For example, since FFD has limited capability for complex shape deformation, our prior method required a dense grid of thousands of control point to achieve acceptable whole heart reconstruction accuracy. Herein we demonstrate that using control-handle-based deformation with biharmonic coordinates achieves higher reconstruction accuracy while using far less control points than the FFD-based approach.

\section{Methods}
\subsection{Shape Deformation Using Biharmonic Coordinates}
We parameterize deformations of whole heart meshes with the translations of a small set of deformation handles sampled from the mesh template. Given a set of mesh vertices $V \in \mathbb{R}^{n\times 3}$ and a set of control points $P \in \mathbb{R}^{c\times 3}$, we compute the biharmonic coordinates $W\in\mathbb{R}^{n\times c}$, which is a linear map, $V = WP$. $n$ and $c$ are the number of vertices and the number of control points, respectively. $W$ is defined based on biharmonic functions and can be pre-computed by minimizing a quadratic deformation energy function while satisfying the handle constraints with linear precision \cite{Wang2015LinearSD}. Namely, let $Q\in \mathbb{R}^{c\times n}$ be the binary selector matrix that selects rows of $X$ corresponding to the control handles, and let $T\in \mathbb{R}^{(n-c)\times n}$ be the complementary selector matrix of $Q$ corresponding to the free vertices. W is computed by
\begin{align}
        V &= \argmin_{X\in\mathbb{R}^{n\times3}} \frac{1}{2} \text{trace}(X^TAX), \quad \text{subject to } QX = P \\
        V &= \underbrace{(Q^T - T^T(TAT^T)^{-1}TAQ^T)}_{W}P
\end{align}
where $A$ is a positive semi-definite quadratic form based on the squared Laplacian energy to encourage smoothness \cite{Wang2015LinearSD}. Under this framework, displacements of the control handles can smoothly deform the underlying mesh template. 

\subsection{Network Architecture}

\begin{figure*}[ht]
    \centering
    \includegraphics[width=\linewidth]{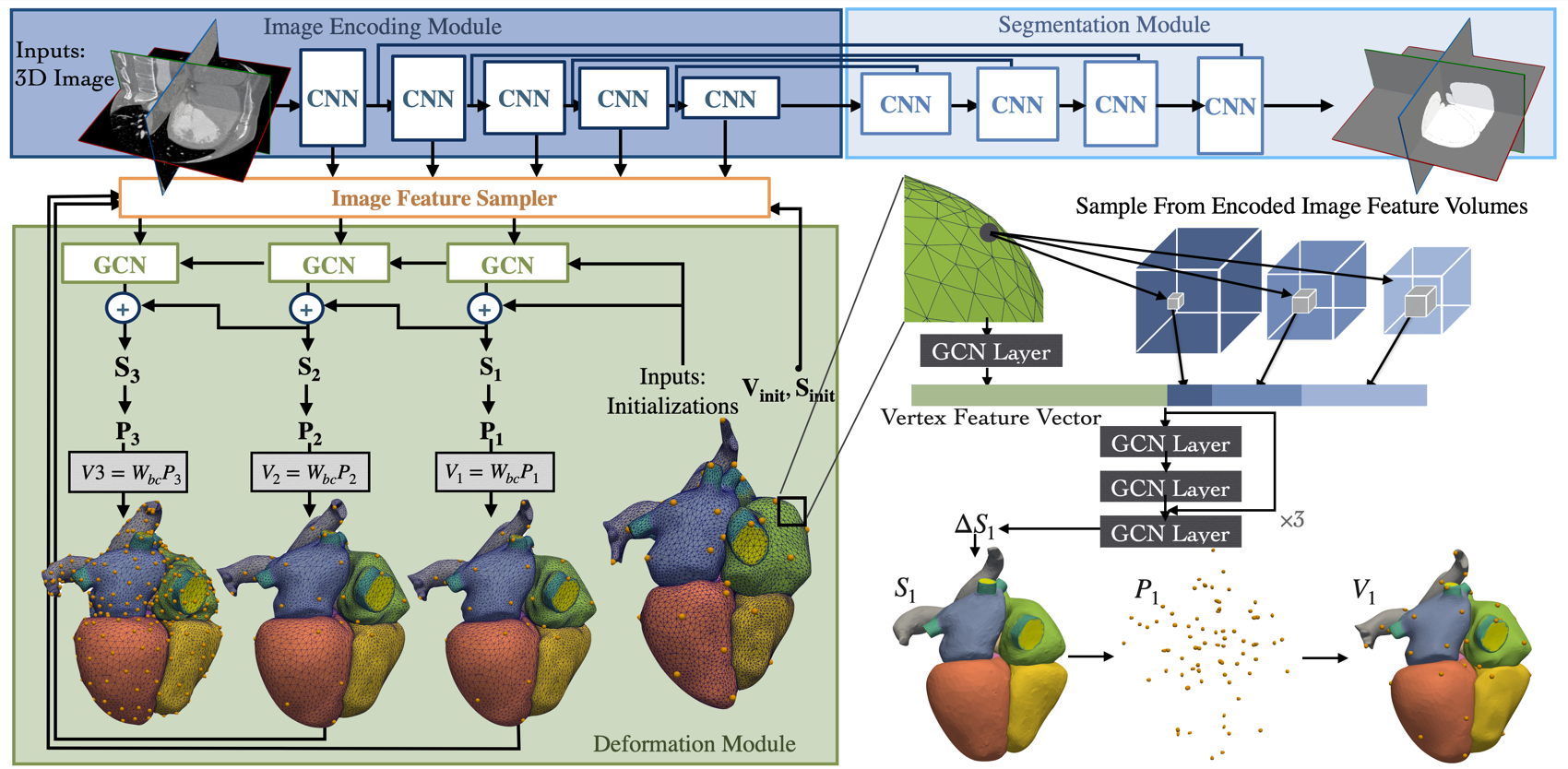}
    \caption{Diagram of the proposed automatic whole heart reconstruction approach. A total of three deformation blocks were used to progressively deform the mesh templates, using 75, 75 and 600 control handles, respectively, for the 3 deformation blocks. }
    \label{fig:network}
\end{figure*}

Figure \ref{fig:network} shows the overall architecture of our network. The central architecture is the novel control-handle-based mesh deformation module, which learns to predict the displacements of control handles based on image features, so that the underlying mesh templates can be smoothly deformed to match the input 3D image data. 

\subsubsection{Image Encoding and Segmentation Modules} We applied a residual 3D CNN backbone to extract and encode image features at multiple resolutions \cite{Isensee2019AnAA}. The CNN backbone involves 4 down-sampling operations so that image feature volumes at 5 different resolution are obtained. These image feature volumes are used as inputs to GCN layers to predict the displacements of control handles. Similar to \cite{kong2021deeplearning}, we also used a segmentation module that predicted a binary segmentation map to enable additional supervision using ground truth annotations. This module was only used during training. 

\subsubsection{Mesh Deformation Module} Biharmonic coordinates constrain the displaced control handles to be located on the deformed mesh template. Therefore, regardless of which set of control handles are sampled, these handles will be located at the corresponding positions on the template mesh. We used a neural network to update the coordinates of all points ($S\in \mathbb{R}^{n\times 3}$) on the mesh template and obtain the coordinates of the selected control handles ($P \in \mathbb{R}^{c\times 3}$) from the updated mesh vertex locations to deform the template using the pre-computed biharmonic coordinates. This design allows picking arbitrary sets of control handles to deform the template at various resolutions after training. Furthermore, training to predict the coordinates of every mesh vertex provides additional supervision that can speed up training.  

Since the mesh template can be represented as a graph, a GCN was used to predict the mesh vertex displacements. We chose to approximate the graph convolutional kernel with a first order Chebyshev polynomial of the normalized graph Laplacian matrix \cite{Defferrard2016}. At each mesh vertex, we extracted the image feature vectors at the corresponding image coordinates from multiple image feature volumes with various resolution. These image feature vectors were then concatenated with the mesh feature vectors following a GCN layer. The combined vertex feature vectors were then processed by three graph residual blocks. We then used an additional GCN layer to predict displacements as 3D feature vectors on mesh vertices. We used a total of three deformation blocks to progressively deform the template mesh. The first and second deformation blocks used 75 control handles to deform the mesh, whereas the last deformation block used more control handles, 600, for a more detailed prediction.

\subsection{Network Training}

\begin{figure}[ht]
    \centering
    \includegraphics[width=\columnwidth]{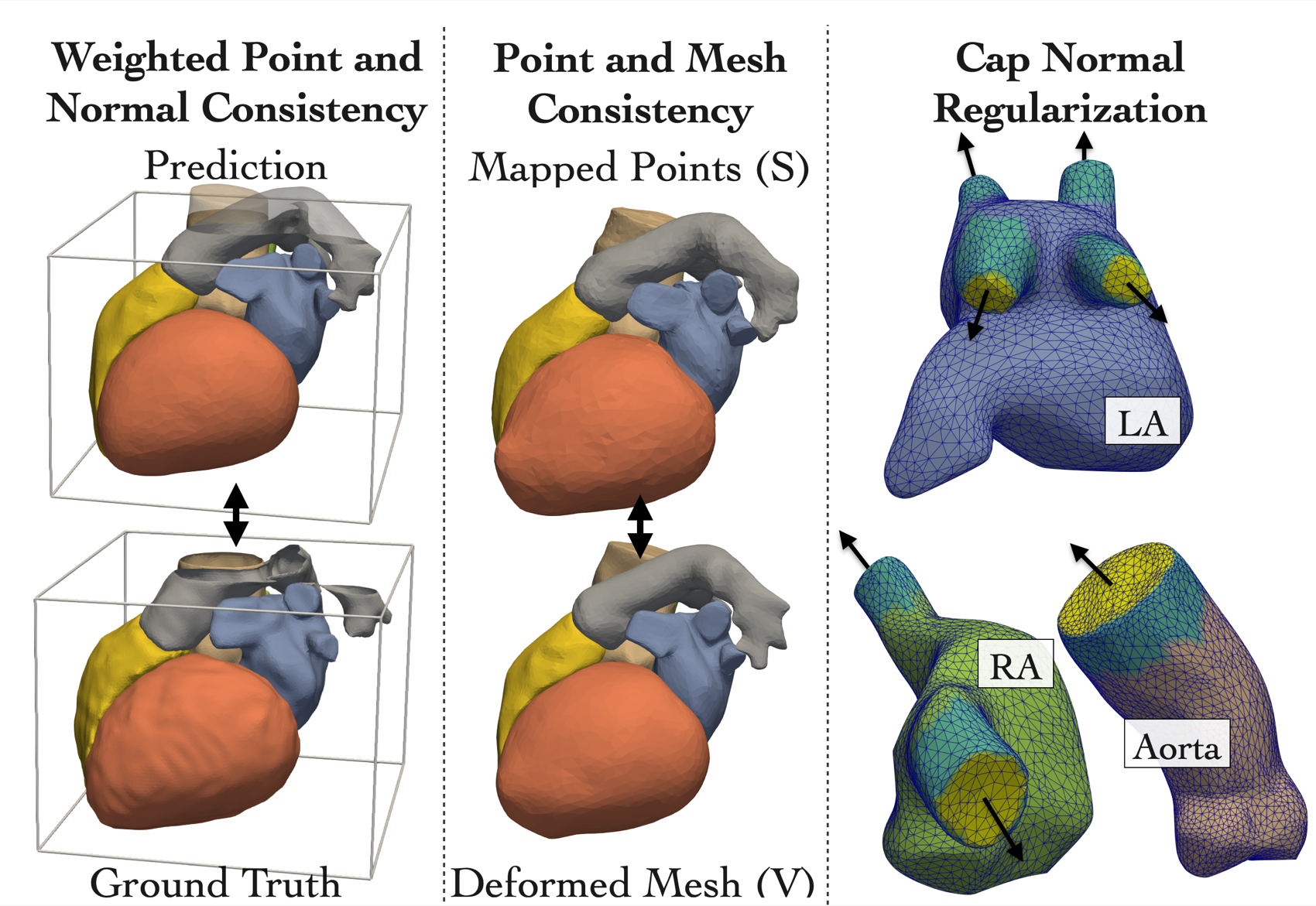}
    \caption{Graphical illustration of different loss functions. Yellow and teal on the right panel shows the caps and walls to apply the mesh regularization losses, respectively, and arrows shows cap normal vectors.}
    \label{fig:method_losses}
\end{figure}
The training of the networks was supervised by 3D ground truth meshes of the whole heart as well as a binary segmentation indicating occupancy of the heart on the input image grid. We used the following loss functions (illustrated in figure \ref{fig:method_losses}) in training to produce accurate whole heart geometries while ensuring the resulting mesh is suitable to support simulations. 

\subsubsection{Geometric Consistency Losses}
The geometric consistency loss $\mathcal{L}_{geo}$ is the geometric mean between the point and normal consistency losses to supervise the geometric consistency between the prediction and the ground truth\cite{kong2021deeplearning}. 
We note that edge length and Laplacian regularization losses, such as used in \cite{kong2021deeplearning}, are not necessary since the smoothness of the mesh prediction is naturally constrained by the biharmonic coordinates used to deform the template. 
Since only the selected control points were used to deform the mesh template while the displacements of all mesh points were predicted, we needed to regularize the $L2$ distances between the mapped mesh points ($S \in \mathbb{R}^{n\times 3}$) and the corresponding mesh vertices on the deformed mesh template ($V \in \mathbb{R}^{n\times 3}$). This consistency loss between the points and the mesh ensures that coordinates of other unselected control points also result in reasonable deformations. In other words, the deformation results should not be sensitive to the choice of pre-selected control points. 

\subsubsection{Mesh Regularization for CFD Simulations}
Cardiac models generally includes portions of the great vessels connected to the heart (e.g., pulmonary veins and arteries, venae cavae, and aorta). For CFD simulation of cardiac flow, locations where these vessels are truncated (so-called inflow and outflow boundaries, or ``caps'') should be planar and nominally orthogonal to the vessel. On our training template, we labeled these caps, as well as the associated vessel walls. Figure \ref{fig:method_losses} shows the identified cap and wall faces on left atrium (LA), right atrium (RA) and aorta. We applied a co-planar loss on each cap that penalizes the $L2$ differences of the surface normals on the cap. Namely, 
$
    \mathcal{L}_{coplanar} = \sum_k \sum_{j\in \mathcal{C}_k} || \mathbf{n}_j - \frac{1}{|C_k|}\sum_{j\in \mathcal{C}_k} \mathbf{n}_j ||_2^2
$
where $C_k$ is the set of mesh faces for the $k$th cap and $\mathbf{n}_j$ is the normal vector for the $j$th face on $C_k$. For mesh vertices that are on the vessel walls near the caps, we minimized the absolute value of the dot product between the surface normal vectors and the surface normal vector of the caps to encourage orthogonality. Namely,
$
    \mathcal{L}_{orthogonal} = \sum_k \sum_{j\in \mathcal{W}_k} | \langle \mathbf{n}_k, \frac{1}{|C_k|}\sum_{p\in \mathcal{C}_k} \mathbf{n}_p \rangle | \;,
$
where $\mathcal{W}_k$ is the set of mesh faces on the vessel wall that corresponds to the $k$th cap.

\subsubsection{Weighted Mesh Losses}
Patient images may not always contain the targeted cardiac structures. As shown in Figure \ref{fig:method_losses} (left), cardiac structures such as pulmonary veins, pulmonary arteries and the aorta are often not captured in full, although the truncks of these structures can be necessary for simulations. We thus aim to predict ``complete'' whole heart structures from incomplete image data. Namely, we computed the bounding box of the ground truth meshes and assigned zero weights within the geometric consistency loss for predicted mesh vertices that were located outside the bounding box. Furthermore, as the geometry of inlet vessels are important to the accuracy of CFD results, we applied a higher weight for the geometric consistency loss on mesh vertices that are located on vessel walls near the inlets. 

\subsubsection{Total Losses}
The total loss on a predicted mesh $M$ is 
\begin{equation}
\begin{split}
    \mathcal{L}_{mesh}(M, &G, W) = \sum_i \mathcal{L}_{geo}(M_i, G_i, \mathbf{w}_i) \\
    & +\alpha \sum_k (\mathcal{L}_{coplanar} (M) + \beta \mathcal{L}_{orthogonal}(M))
\end{split}
\end{equation}
where $G_i$ represents the ground truth mesh for individual cardiac structure, and $\mathbf{w}_i$ represents the weighting vector for each mesh point. $M$ can be both the deformed whole heart mesh template $V$ and the mesh obtained from mapping all mesh points $S$. The total loss for training is a weighted sum of the mesh losses and the segmentation loss, which is the sum of the binary cross-entropy and the dice losses between the predicted occupancy probability map $I_p$ and the ground truth binary segmentation of the whole heart $I_g$.
\begin{equation}
\begin{split}
    \mathcal{L}_{total} &= \lambda_1 \mathcal{L}_{mesh}(S, G, W) + \lambda_2 \mathcal{L}_{mesh}(V, G, W) + \\
    & \lambda_3||S - V||_F^2 + \mathcal{L}_{seg}(I_g, I_p)
    \end{split}
\end{equation}

\subsubsection{Image Augmentation for Shape Completion}
\begin{figure}[ht]
    \centering
    \includegraphics[width=\columnwidth]{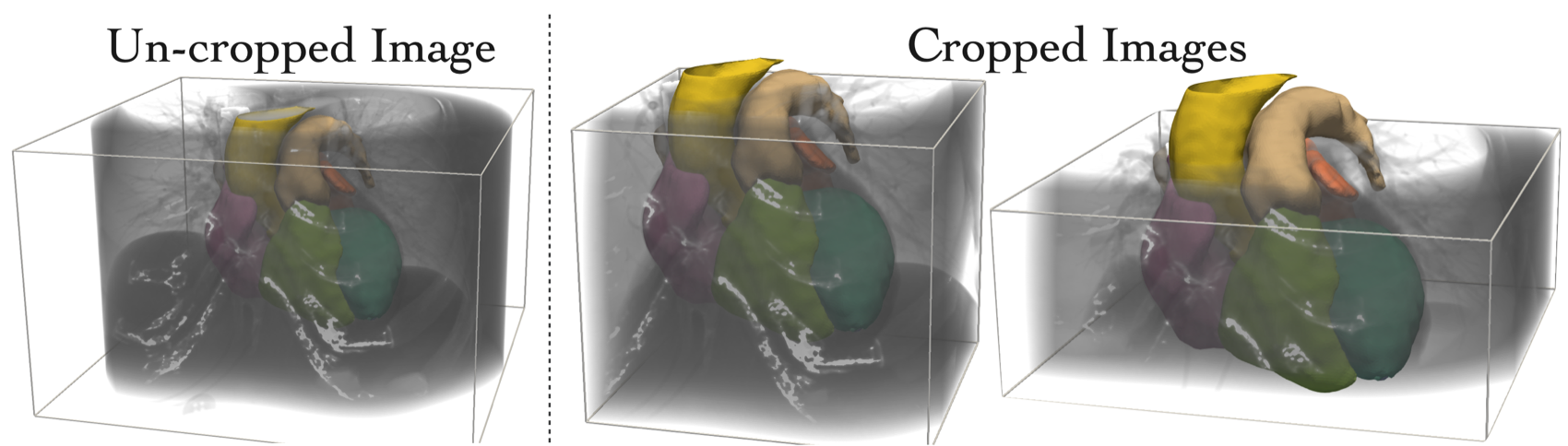}
    \caption{Visualization of augmented input image crops and the corresponding ground truth meshes. }
    \label{fig:method_crop}
\end{figure}

Leveraging the mesh template, we trained our method to predict the geometries of the whole heart represented by the template mesh when the images do not cover the complete cardiac structures. Since CT images often do not cover the whole heart, we selected CT images that did cover the whole heart (n=10) from our training set and then generated 10 random crops for each image while keeping the ground truth meshes to be the same. Figure \ref{fig:method_crop} visualizes example image crops and the corresponding ground truth meshes.
We also applied random shearing, rotations and elastic deformations, following the same augmentation strategies described in our prior work \cite{kong2021deeplearning}.

\section{Experiments}
\subsection{Datasets and Experiments} 
\subsubsection{Task-1: Whole Heart Segmentation for 3D Images} 
\label{sec:task1}
We applied our method to public datasets of contrast-enhanced 3D CT images and 3D MR images from both normal and abnormal hearts. For training and validation, we used a total of 102 CT images and 47 MR images from the multi-modality whole heart segmentation challenge (MMWHS) \cite{ZHUANG2019}, orCalScore challenge\cite{orCaScore}, left atrial wall thickness challenge \cite{SLAWT} and left atrial segmentation challenge \cite{LASC}. Among them, we used 87 CT and 41 MR images for training, and we used 15 CT images and 6 MR images for validation, where we tuned the hyper-parameters and selected the model trained with the hyper-parameter set that performed best on the validation data. We then evaluated the final performance of the selected model on a held out test dataset from the MMWHS challenge, which contained 40 CT and 40 MR images. For CT images, the in-plane resolutions vary from $0.4 \times 0.4$ mm to $0.78 \times 0.78$ mm and the through-plane resolutions vary from $0.5$ mm to $1.6 $ mm. For MR images, the in-plane resolutions vary from $1.25 \times 1.25$ mm to $2 \times 2.$ mm and the through-plane resolutions vary from $2.$ mm to $2.3$ mm.

\begin{figure}[ht]
    \centering
    \includegraphics[width=\columnwidth]{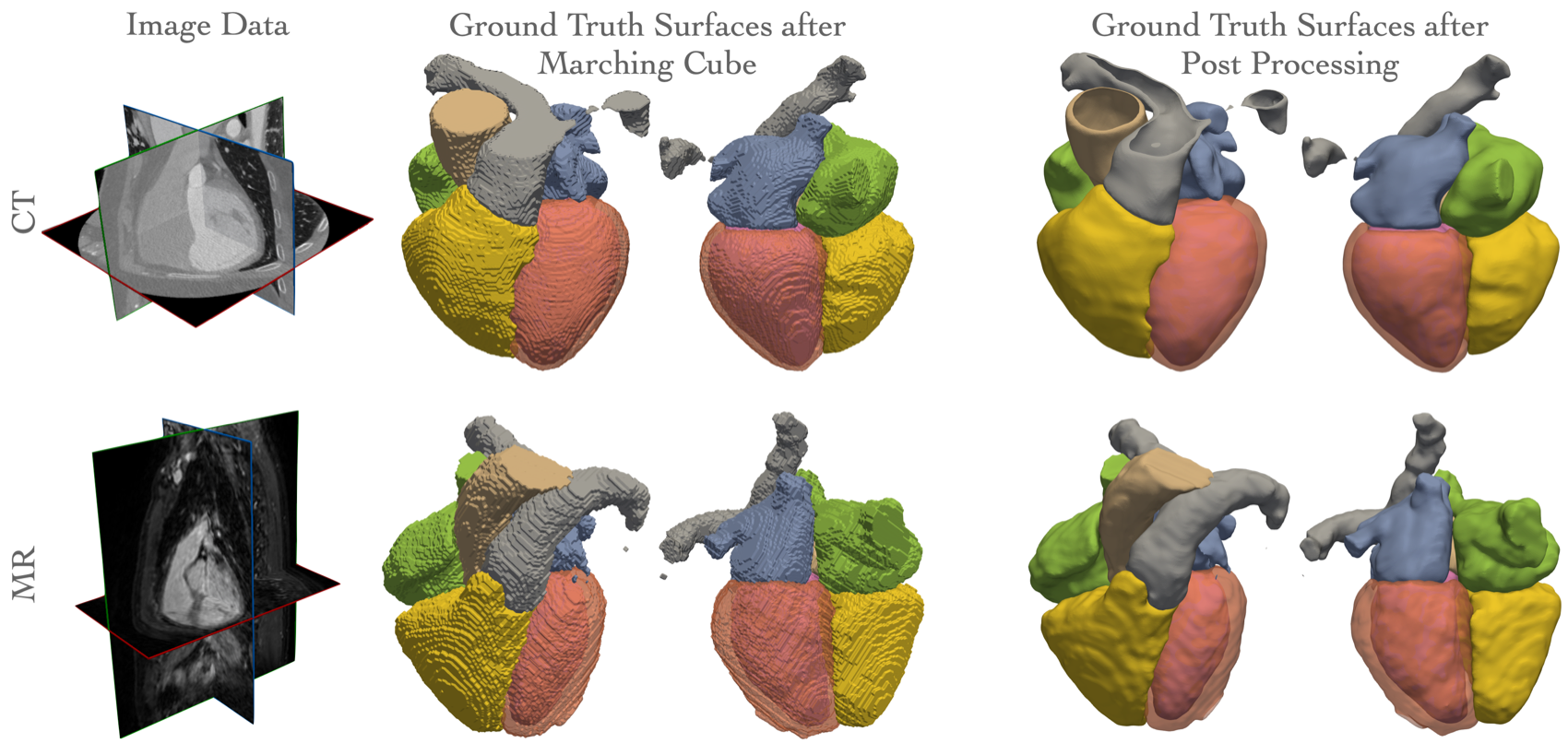}
    \caption{Illustration of example CT and MR image data, the corresponding surface meshes generated from manual segmentation using the marching cube algorithm, and the resulting ground truth surface meshes after post processing.}
    \label{fig:gt_mesh}
\end{figure}

For each image in the dataset, we followed the MMWHS challenge \cite{ZHUANG2019} and created ground truth segmentation of 7 cardiac structures to supervise the training and evaluate model performance on validation and test datasets. The 7 cardiac structures included the blood cavities of left ventricle (LV), right ventricle (RV), left atrium (LA), right atrium (RA), LV myocardium (Myo), aorta (Ao), and pulmonary artery (PA), for all images. Figure \ref{fig:gt_mesh} illustrates our pipeline to generate smooth ground truth surface meshes from the manual segmentations. We resampled the segmentation to a resolution of $1. \times 1. \times 1.$ mm, then used the marching cube algorithm to generate the surface meshes for each cardiac structure. We then applied a Windowed-Sinc smoothing filter \cite{Taubin1996OptimalSS} with a low pass band of 0.01 and 20 iterations of smoothing to generate smooth ground truth meshes. Furthermore, as visualized in the example CT case in Figure \ref{fig:gt_mesh}, surface meshes were clipped at the image bounding box to remove the fictitious surface at the image boundaries for cardiac structures that exceeded the coverage of the image data.

We compared the geometric accuracy of the reconstructed whole heart surfaces against prior deep-learning methods that demonstrated strong performance of segmenting whole heart geometries from 3D medical images. Namely, we considered HeartFFDNet\cite{KONG2021FFD}, our prior work that generates simulation-ready whole heart surface meshes from images by learning free-form deformation from a template mesh, MeshDeformNet \cite{kong2021deeplearning} that predicts displacements on sphere mesh templates, as well as 2D UNet \cite{Ronneberger2015} and a residual 3D UNet \cite{Isensee2019AnAA} that are arguably the most successful neural network architecture for image segmentation. We also implemented a SpatialConfiguration Net (SCN) \cite{Payer2018} that ranked first for CT and second for MRI in the MMWHS challenge, using our residual 3D UNet backbone. This segmentation-based approach incorporates relative positions among structures to focus on anatomically feasible regions. All methods were trained on the same training and validation data splits,  and used the same pre-processing and augmentation procedures to ensure a fair comparison. 

\subsubsection{Task-2: Whole Heart Mesh Construction for 4D Images} We applied our method on time-series CT images to evaluate its performance on creating whole heart meshes for CFD simulations. Since the MMWHS dataset does not include pulmonary veins, LA appendage or venae cavae, we prepared another set of ground truth segmentations to include these structures. The geometric accuracy and the mesh quality of the reconstructed meshes for CFD simulations were then evaluated on 10 sets of time-series CT images against the learning-based mesh reconstruction baselines, HeartFFDNet and MeshDeformNet.

\subsubsection{Task-3: CFD Simulations} 
\label{section:cfd_experiments}
\begin{figure}[ht]
    \centering
    \includegraphics[width=0.8\columnwidth]{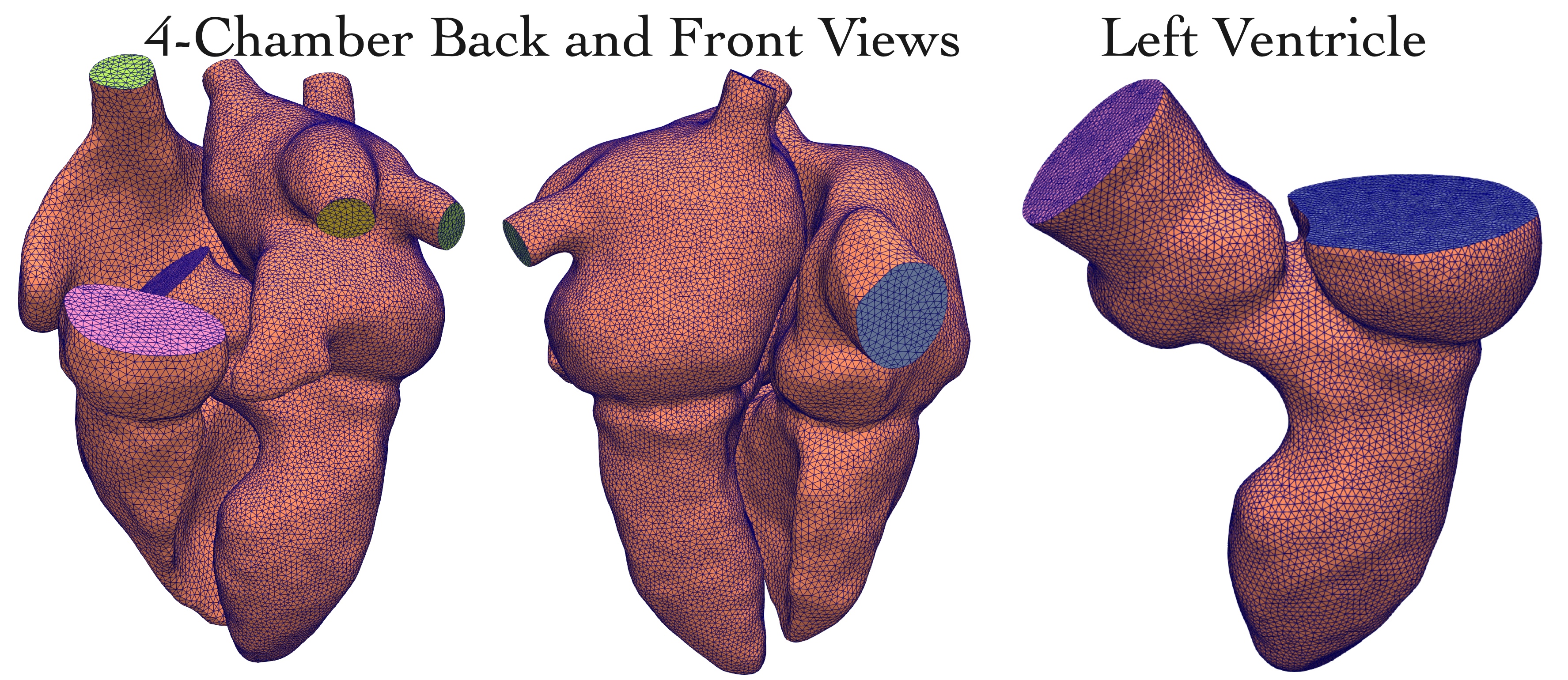}
    \caption{Visualization of simulation-ready templates with trimmed inlet/outlet geometries and tagged face IDs for prescribing boundary conditions.}
    \label{fig:cfd_template}
\end{figure}
We conducted CFD simulations of cardiac flow using the predicted whole heart meshes from time-series CT images. Since our predicted model does not contain heart valves, only diastolic flow was simulated. We also conducted CFD simulations for the LV and simulated the LV flow for the entire cardiac cycle. Figure \ref{fig:cfd_template} visualizes the simulation-ready templates of the 4 heart cambers and the LV with trimmed inlet/outlet geometries and tagged face IDs for prescribing boundary conditions. These simulation-ready templates were manually created from the training whole heart template in a surface processing software, SimVascular \cite{Updegrove2016}.We linearly interpolated the pre-computed biharmonic coordinates onto the new templates so that our trained models can readily deform these new templates. The simulation results were compared against results obtained from time-series ground truth meshes created manually in SimVascular \cite{Updegrove2016}. We also compared simulation results among our method, HeartFFDNet, and a conventional semi-automatic model construction pipeline based on image registration, where a manually created ground truth mesh was morphed based on transformations obtained from registering images across different time points.

\subsection{Evaluation Metrics} 
We used Dice similarity coefficient (DSC) and Hausdorff Distance (HD) to measure segmentation accuracy. The DSC and HD values for the MMWHS test dataset were evaluated with an executable provided by MMWHS organizers. For mesh-based methods, we converted the predicted surface meshes to segmentation prior to evaluation. Mesh quality was compared in terms of the percentage self-intersection, which measures the local topological correctness of the meshes, orthogonality of the vessel caps with respect to the vessel walls, as well as the coplanarity of the vessel caps. The percentage mesh self-intersection was calculated as the percentage of intersected mesh facets detected by TetGen \cite{Si2015} among all mesh facets. The orthogonality between vessel caps and walls (Cap-Wall-Orthogonality) was measured by the normal consistency between the mean cap normal vector and the vector connecting the centroids of the mesh points on the cap and on the wall, respectively. Namely,
$
    CWO = \sum_{k} 1 - \langle \frac{1}{|\mathcal{W}_k|} \sum_{i\in\mathcal{W}_k|} \mathbf{n}_i, \frac{1}{|\mathcal{C}_k|} \sum_{j\in\mathcal{C}_k|} \mathbf{n}_j \rangle
$
where $\mathcal{W}_k$ and $\mathcal{C}_k$ represent the sets containing the mesh vertices on a vessel wall and the corresponding vessel cap. Vessel caps coplanarity was measured by the projected distance between the mesh vertices on the cap and the best fit plane over those mesh vertices. For CFD simulations, we compared integrative measures during a cardiac cycle, namely, LV volume and average kinetic energy $KE' = \frac{1}{2V_{LV}} \int \int \int \rho u^2 dV$, where $V_{LV}$ is the volume of the LV and $u$ is the flow velocity. We also compared the mean velocity near the mitral valve opening (MO) and aortic valve opening (AO) during a cardiac cycle. Paired t-test was used for statistical significance.

\subsection{Deforming-Domain CFD simulations of Cardiac Flow}
We applied the Arbitrary Lagrangian-Eulerian (ALE) formulation of the incompressible Navier-Stokes equations to simulate the intraventricular flow and account for deforming volumetric mesh using the finite element method. Since time resolution of image data is too coarse (about 0.1s) to be used directly in time-stepping of the Navier–Stokes equations, cubic spline interpolation was applied to interpolate the meshes predicted at different imaging time points so that the time resolution of the interpolated meshes was 0.001s, which corresponded to the simulation time step. For the fluid domain, the mesh motions computed from these interpolated meshes were imposed as Dirichlet boundary conditions on the chamber walls. For simulations of LV flow, we imposed Dirichlet boundary conditions on the mitral inlet during systole, and on the aortic outlet during diastole. Neumann (prescribed pressure) boundary conditions were applied to the mitral inlet during diastole or to the aortic inlet during systole. Diastole and systole phases were determined based on the increase and decrease of the LV volume. For simulations of diastolic cardiac flow within 4 heart chambers, we applied Neumann boundary conditions to the pulmonary vein inlets, and imposed Dirichlet boudary condition on the aortic outlet. Blood was assumed to have a viscosity $\mu$ of $4.0 \times 10^{-3} Pa \cdot s$ and a density $\rho$ of $1.06 g/cm^3$. The volumetric mesh was created automatically from our predicted surface mesh using TetGen \cite{Si2015}, using a maximum edge size of 1.5mm. The equations were solved with the open-source svFSI solver from the SimVascular project \cite{ChisvFSI}.

\section{Results}
\subsection{Comparative Studies of Whole Heart Segmentation Performance on MMWHS Dataset}
\subsubsection{Comparison of Geometric Accuracy with Other Methods}

\begin{table*}[ht]
\caption{Comparison of Whole-Heart Segmentation Performance, DSC ($\uparrow$)and HD (mm) ($\downarrow$), from different methods on the MMWHS CT and MR test datasets.* Denotes Significant Difference Of "Ours (S)" From the Others (p-Values<0.05)}
\label{table:mmwhs}
\resizebox{\textwidth}{!}{%
\begin{tabular}{llllllllll|lllllllll}
\toprule
&        &               &               &                &               CT &                &                &                &                &               &                &                &               MR &                &                &                &               \\
\midrule
&  Method      &               Myo &               LA &               LV &               RA &               RV &               Ao &               PA &               WH &              Myo &               LA &               LV &               RA &               RV &               Ao &               PA &               WH\\
\midrule
\multirow{7}{*}{DCS} 
            & Ours (S) &            90.07 &            93.18 &            93.47 &   \textbf{89.48} &   \textbf{91.48} &            93.33 &   \textbf{85.60} &            91.76 &   \textbf{80.45} &            86.98 &            91.61 &            88.08 &            88.09 &            85.76 &            78.14 &            87.41\\
            & Ours (V) &           88.38* &           92.53* &           91.99* &           88.76* &           90.59* &           91.25* &           84.73* &           90.53* &           78.62* &           86.27* &           89.38* &           87.79* &           87.20* &           83.30* &            77.55 &           86.04*\\
            & HeartFFDNet &           83.85* &           90.55* &           89.38* &           86.33* &           87.65* &           90.65* &           80.20* &           87.82* &           70.67* &           83.27* &           86.92* &           84.47* &           82.77* &           79.71* &           69.68* &           81.33*\\
            & MeshDeformNet &            89.94 &   \textbf{93.23} &  \textbf{93.98}* &            89.18 &            91.00 &  \textbf{94.98}* &            85.22 &   \textbf{91.80} &            79.71 &   \textbf{88.13} &   \textbf{92.23} &   \textbf{88.82} &  \textbf{89.24}* &  \textbf{88.98}* &  \textbf{81.65}* &  \textbf{88.17}*\\
            & 2DUNet &            89.87 &            93.08 &            93.06 &           87.71* &           90.49* &            93.43 &           83.23* &           91.09* &            79.47 &            86.41 &           89.61* &            85.21 &           86.48* &            86.94 &            77.24 &           85.94*\\
            & 3DUNet &           86.34* &           90.17* &           92.28* &           86.77* &           87.58* &           92.34* &           81.29* &           88.78* &           76.11* &           85.20* &           87.90* &           86.63* &           82.77* &           74.18* &            76.38 &           84.04*\\
            & 3DUNet+SCN &   \textbf{90.09} &           92.63* &            93.43 &            89.01 &           90.49* &            91.61 &            84.05 &           91.18* &           78.36* &           85.77* &           89.84* &            88.09 &            87.19 &            84.35 &            76.85 &           86.12*\\
\cline{1-18}
    \multirow{7}{*}{HD} 
            & Ours (S) &            14.41 &            10.72 &            10.41 &            13.80 &            11.63 &             6.59 &             7.88 &            16.95 &            16.39 &            12.12 &            11.93 &            13.93 &            14.76 &             7.19 &             9.12 &            19.97\\
            & Ours (V) &            14.40 &   \textbf{8.18}* &            6.87* &  \textbf{12.46}* &   \textbf{9.55}* &            5.54* &             8.45 &           16.63* &  \textbf{15.96}* &  \textbf{10.16}* &   \textbf{8.97}* &  \textbf{12.56}* &  \textbf{12.46}* &             7.39 &             9.14 &  \textbf{18.91}*\\
            & HeartFFDNet &            14.20 &            8.74* &            7.66* &            13.24 &           10.44* &             6.17 &             8.35 &            16.57 &           18.21* &            12.43 &            12.57 &            15.57 &            16.36 &            9.26* &           11.36* &           22.28*\\
            & MeshDeformNet &            14.39 &            10.41 &            10.33 &            13.67 &            13.36 &   \textbf{5.27}* &            9.16* &            17.62 &            16.92 &            12.22 &            11.63 &            15.05 &            14.73 &   \textbf{6.05}* &   \textbf{7.79}* &            21.08\\
            & 2DUNet &   \textbf{9.98}* &             9.34 &   \textbf{6.10}* &            13.78 &            10.39 &             5.63 &             8.38 &   \textbf{16.37} &            20.34 &           10.78* &            10.62 &            17.43 &            16.32 &             6.98 &             8.09 &           27.54*\\
            & 3DUNet &            13.64 &            11.47 &            10.12 &           16.56* &           16.17* &             6.88 &            9.88* &           19.91* &           34.97* &           33.81* &           36.28* &           24.36* &           27.72* &           15.52* &            10.13 &           51.54*\\
            & 3DUNet+SCN &            13.78 &            12.80 &             9.86 &            17.02 &           17.70* &             6.57 &    \textbf{7.77} &           28.04* &           39.30* &           34.87* &           21.06* &           28.37* &           28.20* &             8.05 &             8.49 &           53.76* \\
\bottomrule
\end{tabular}
}
\end{table*}

\begin{figure*}[ht]
    \centering
    \includegraphics[width=\textwidth]{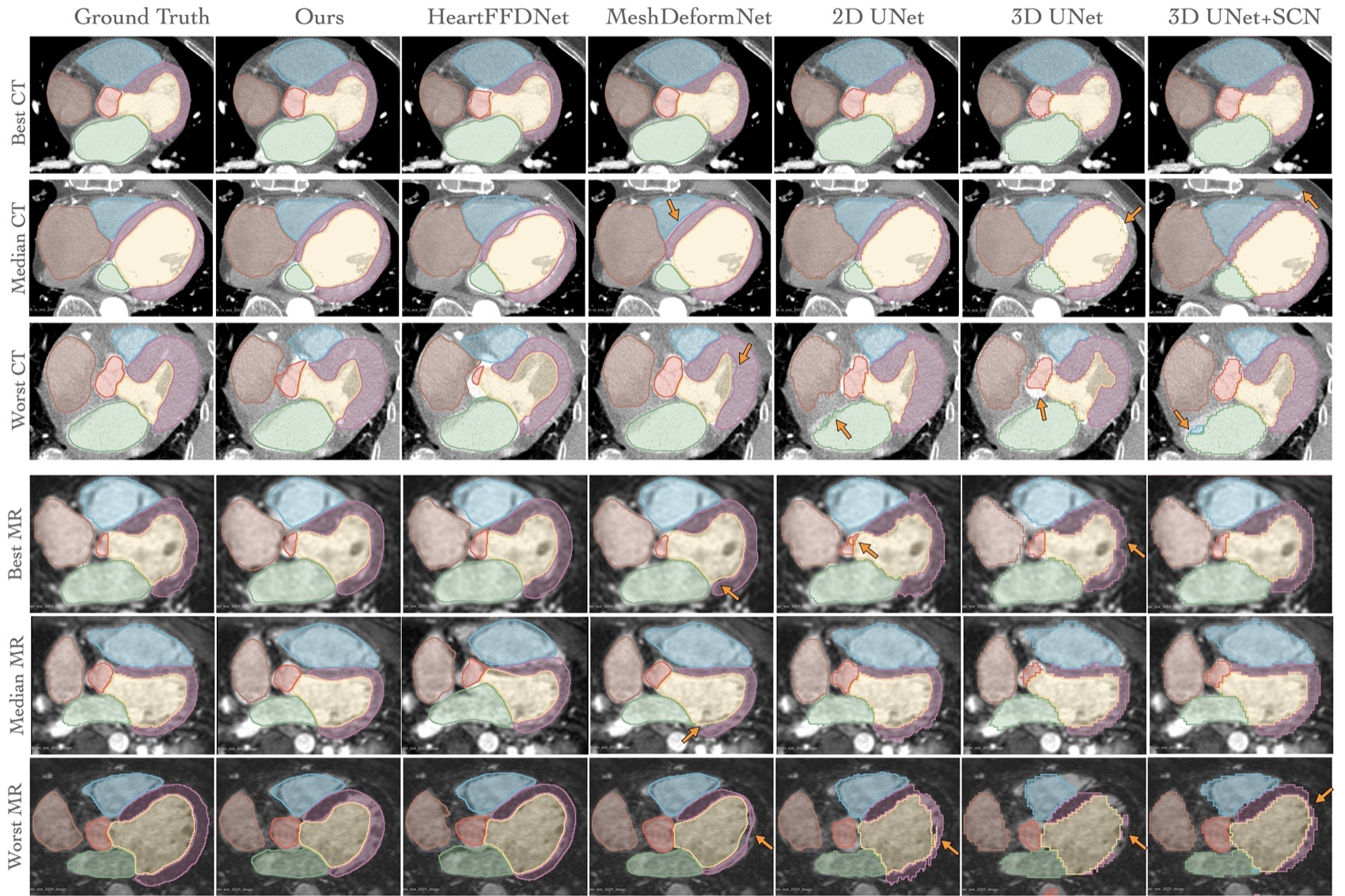}
    \caption{Example segmentation results for CT and MR images from different methods. The CT or MR images that our method had the best, the median, and the worst Dice scores among the CT or MR test data were selected, thus illustrating best, typical, and worst segmentation results, respectively. The gold arrows indicate locations of artifacts unsuitable for simulations, such as gaps between adjacent cardiac structures for MeshDeformNet, missing structures, holes, noisy boundaries, and isolated islands for the segmentation-based methods.}
    \label{fig:example_seg}
\end{figure*}

Table \ref{table:mmwhs} compares the average Dice scores and Hausdorff distances of the reconstruction results of both the whole heart and the individual cardiac structures for the MMWHS test dataset. We show the accuracy of deforming the template by mapping all mesh points ($S$) and by interpolating the mesh deformation using only 600 uniformly-sampled control handles ($V$). Mapping all points consistently achieved higher dice scores than using 600 selected control handles, but the HDs are worse for some cardiac structures. For both CT and MR data, in terms of Dice scores, our method consistently outperformed HeartFFDNet and 3D UNet for all cardiac structures and achieved comparable performance with MeshDeformNet, 2D UNet, 3D SCNet for most cardiac structures. Our method achieved the best HDs for LA, RA and RV for CT data and for all cardiac structures except for aorta and PA for MR data.

Figure \ref{fig:example_seg} presents the best, median, and worst segmentation results of our method on CT and MRI test images and provides qualitative comparisons of the results from the different methods. As shown, mesh-based approaches, ours, HeartFFDNet and MeshDeformNet produced smooth and anatomically consistent cardiac geometries while segmentation-based approaches, 2D UNet, residual 3D UNet, 3D SCNet produced segmentations with topological artifacts such as missing parts, holes, and isolated islands. Although 3D SCNet produced higher Dice scores than residual 3D UNet, it produced a few misclassifications where the LA was incorrectly classified into RV. Although MeshDeformNet produced smooth and anatomically consistent cardiac geometries, it was prone to gaps between adjacent cardiac structures by deforming un-coupled spheres. Our method and the HeartFFDNet were able to avoid this limitation by deforming the space enclosing a whole heart template, preserving the connections among cardiac structures. 

\subsubsection{Effect of Varying Control Handle Numbers}
We investigated the effect of various design choices on the whole heart segmentation performance of our proposed method. Table \ref{table:control_handle_numbers}, presents the effect of varying the number of control handles used during training on the average Dice scores and Hausdorff distances of the reconstruction results. Increasing the number of control handles used in the last deformation block from 75 to 900 generally resulted in increased performance for most cardiac structures. However, the resulting improvement in terms of Dice scores was only around $1\%$, indicating the robustness of our method towards using relatively fewer numbers of control handles. Similarly, using more control handles in the first and/or second deformation blocks did not result in significant improvement for most cardiac structures. Therefore, in our final network model, we chose to use a small number of control handles (75) in the first and second blocks to reduce the computational cost, and used 600 control handles in the last deformation block for a slightly better performance.

\begin{table*}[ht]
    \centering
    \caption{A comparison of  Whole-Heart Segmentation Performance, DSC ($\uparrow$)and HD (mm) ($\downarrow$) on the MMWHS CT and MR test datasets, When Using Different Numbers of Control Handles During TraIning. B1/B2/B3 Denotes the Numbers of Control Handles Used in the Three Deformation Blocks. * Denotes Significant Difference Of Our Final Network Model using "75/75/600" Control Handles From the Others (p-Values<0.05)}
    \label{table:control_handle_numbers}
    \resizebox{\textwidth}{!}{%
\begin{tabular}{llllllllll|lllllllll}
\toprule
&        &               &               &                &               CT &                &                &                &                &               &                &                &               MR &                &                &                &               \\
\midrule
& B1/B2/B3        &               Myo &               LA &               LV &               RA &               RV &               Ao &               PA &               WH &              Myo &               LA &               LV &               RA &               RV &               Ao &               PA &               WH\\
\midrule
        \multirow{7}{*}{DCS} 
                    & 75/75/600 &  \textbf{90.07} &           93.18 &   \textbf{93.47} &           89.48 &            91.48 &            93.33 &            85.60 &  \textbf{91.76} &  \textbf{80.45} &           86.98 &            91.61 &           88.08 &   \textbf{88.09} &            85.76 &            78.14 &           87.41 \\
            & 75/75/75 &          88.08* &          92.48* &           92.54* &          88.42* &           90.98* &            92.76 &            84.77 &          90.77* &           79.51 &           86.57 &           90.71* &           87.42 &            87.55 &           83.21* &           74.40* &          86.33* \\
            & 75/75/150 &          88.37* &          92.59* &           92.37* &          88.26* &           90.65* &            92.78 &           84.20* &          90.69* &          78.40* &          85.84* &           90.64* &          86.99* &           86.84* &           82.81* &           74.67* &          85.89*\\
            & 75/75/300 &          88.31* &  \textbf{93.47} &            93.28 &           89.20 &            91.28 &            93.96 &            84.84 &          91.34* &           80.23 &  \textbf{87.98} &  \textbf{92.39}* &          87.08* &            87.94 &   \textbf{86.84} &   \textbf{78.98} &  \textbf{87.46}\\
            & 75/75/900 &          89.04* &           93.20 &            93.36 &  \textbf{89.54} &   \textbf{91.64} &  \textbf{93.97}* &   \textbf{85.99} &           91.66 &          78.64* &           87.07 &            91.54 &          86.91* &            86.77 &            84.73 &            75.22 &          86.48*\\
            & 75/300/600 &           89.91 &           93.22 &            93.27 &           89.17 &           90.73* &            93.83 &            85.48 &           91.52  &           79.74 &          86.00* &            91.18 &  \textbf{88.26} &            87.61 &            83.94 &           73.17* &          86.66*\\
            & 600/600/600 &           89.53 &           93.01 &            93.07 &          88.43* &           90.87* &            93.32 &            84.95 &          91.25* &           79.87 &           86.32 &            91.33 &          87.20* &            87.34 &           83.42* &           73.59* &          86.55*\\
        \cline{1-18}
           \multirow{7}{*}{HD} 
            & 75/75/600 &           14.41 &           10.72 &            10.41 &           13.80 &            11.63 &             6.59 &            7.88 &           16.95 &           16.39 &           12.12 &            11.93 &           13.93 &            14.76 &             7.19 &            9.12 &  \textbf{19.97}\\
            & 75/75/75 &           14.33 &           10.89 &            11.18 &           14.14 &   \textbf{11.21} &             6.87 &            7.88 &  \textbf{16.76} &  \textbf{16.24} &           12.19 &            12.43 &           14.62 &            14.18 &             8.22 &            9.90 &           20.00\\
            & 75/75/150 &           14.33 &   \textbf{9.98} &    \textbf{9.75} &          15.38* &           12.33* &            5.86* &           9.40* &           17.59 &           16.54 &           12.33 &   \textbf{11.89} &          15.27* &            15.31 &            8.48* &          10.86* &           20.53\\
            & 75/75/300 &  \textbf{14.29} &           10.88 &            10.67 &           14.70 &            11.64 &             6.35 &   \textbf{7.80} &           17.74 &           16.95 &           12.45 &            12.78 &          15.44* &           13.55* &    \textbf{7.03} &   \textbf{8.77} &          21.45*\\
            & 75/75/900 &           14.61 &           11.02 &            11.01 &  \textbf{13.44} &            11.64 &             6.30 &           8.93* &           16.84 &          17.93* &           12.88 &            12.94 &           14.19 &            13.99 &             8.12 &          10.50* &          21.37*\\
            & 75/300/600 &           14.50 &           11.20 &            10.90 &           13.79 &            11.29 &   \textbf{5.82}* &            8.32 &           17.36 &           16.63 &  \textbf{12.00} &            12.03 &  \textbf{13.53} &  \textbf{12.30}* &             8.17 &          10.46* &           20.38 \\
            & 600/600/600 &           14.36 &           10.60 &            10.28 &           14.58 &            12.06 &             7.04 &            8.16 &           17.42  &           17.06 &          12.92* &           13.60* &           14.42 &            14.07 &             8.33 &          10.25* &           20.70\\
        \bottomrule
        \end{tabular}
        }
\end{table*}

\subsubsection{Effect of Individual Loss Components on Whole Heart Segmentation Performance}
Since our training pipeline involves a joint supervision of multiple objectives, we performed an ablation study on the total training loss $\mathcal{L}_{total}$ to evaluate the contribution of individual loss components. Namely, we trained network models while removing the segmentation loss $\mathcal{L}_{seg}$, and the $L2$ consistency loss $||S-V||_F^2$ to investigate the effectiveness of supervising a segmentation branch, and the effectiveness of encouraging the consistency between directly mapped mesh points ($S$) and deformed mesh vertex locations ($V$), respectively. We trained two additional models while removing supervision on $V$ or on $S$, respectively, to validate the effectiveness of supervising both meshes together. Table \ref{table:ablation} presents the effect of removing individual loss components on the whole heart segmentation performance evaluated on the MMWHS test dataset. Removing any of the aforementioned objectives resulted in a significant decrease in whole heart Dice scores for both CT and MR data, as well as decreased Dice scores for most cardiac structures. The Hausdorff distances increased significantly for most cardiac structures when supervision on the smoothly deformed mesh template $V$ was removed. However, there were no significant changes in Hausdorff distances for most cardiac structures following the removal of other objectives.

\begin{table*}[ht]
    \centering
    \caption{Impact of individual loss components of $\mathcal{L}_{total}$ on the prediction accuracy on MMWHS MR and CT test datasets. * Denotes Significant Difference Of our final network model "Ours (S)" From the Others (p-Values<0.05)}
    \label{table:ablation}
    \resizebox{\textwidth}{!}{%
\begin{tabular}{llllllllll|lllllllll}
\toprule
&        &               &               &                &               CT &                &                &                &                &               &                &                &               MR &                &                &                &               \\
\midrule
& Models        &               Myo &               LA &               LV &               RA &               RV &               Ao &               PA &               WH &              Myo &               LA &               LV &               RA &               RV &               Ao &               PA &               WH\\
\midrule
        \multirow{5}{*}{DCS} 
                    & Ours (S) &  \textbf{90.07} &           93.18 &  \textbf{93.47} &  \textbf{89.48} &  \textbf{91.48} &           93.33 &  \textbf{85.60} &  \textbf{91.76}  &  \textbf{80.45} &           86.98 &           91.61 &  \textbf{88.08} &  \textbf{88.09} &  \textbf{85.76} &  \textbf{78.14} &  \textbf{87.41}\\
            & w/o segmentation &          87.11* &  \textbf{93.32} &          92.32* &           88.74 &          90.65* &           92.90 &           85.09 &          90.68*  &          77.96* &           86.31 &          90.81* &           87.48 &          87.03* &           85.47 &           77.28 &          86.28*\\
            & w/o L2 &          88.85* &           92.94 &          92.57* &          88.90* &          90.96* &  \textbf{93.68} &          83.87* &          91.10* &          79.57* &           85.96 &          90.56* &          86.75* &           87.01 &           84.46 &          75.00* &          86.23*\\
            & w/o L2+V &          85.89* &           93.18 &          92.34* &           89.11 &          90.83* &           93.41 &           85.38 &          90.57* &          79.08* &           86.92 &  \textbf{91.93} &          87.51* &           87.01 &           85.02 &          75.30* &          86.66*\\
            & w/o L2+S &          86.12* &          92.73* &          90.84* &          88.78* &          90.41* &          91.99* &          83.81* &          90.05* &          78.11* &  \textbf{87.09} &          90.94* &           87.63 &           87.34 &           85.04 &           76.77 &          86.58*\\

        \cline{1-18}
           \multirow{5}{*}{HD} 
            & Ours (S) &           14.41 &           10.72 &           10.41 &  \textbf{13.80} &  \textbf{11.63} &            6.59 &            7.88 &  \textbf{16.95} &  \textbf{16.39} &           12.12 &           11.93 &  \textbf{13.93} &           14.76 &   \textbf{7.19} &            9.12 &  \textbf{19.97} \\
            & w/o segmentation &           14.41 &   \textbf{9.98} &           10.44 &          15.04* &           12.06 &            6.80 &            8.44 &           17.50 &           17.62 &           12.71 &          13.60* &           14.57 &           15.70 &            7.89 &   \textbf{8.91} &           20.73\\
            & w/o L2 &           14.32 &           10.86 &           10.49 &          14.63* &           12.39 &            6.34 &            8.38 &           17.26 &           16.54 &  \textbf{11.84} &  \textbf{11.46} &           14.74 &           14.75 &           8.17* &            9.86 &           20.52 \\
            & w/o L2+V &           14.42 &           10.41 &  \textbf{10.27} &           14.43 &           12.34 &   \textbf{6.30} &  \textbf{6.78}* &           17.55 &           17.09 &           12.47 &           11.79 &           14.40 &  \textbf{14.39} &            7.51 &            9.93 &          21.19*\\
            & w/o L2+S &  \textbf{14.17} &          12.20* &          12.58* &          16.31* &          13.54* &           8.17* &           9.46* &           17.82 &           16.65 &          13.72* &          14.35* &          16.19* &          16.07* &           8.91* &          10.37* &           20.74\\
        \bottomrule
        \end{tabular}
        }
\end{table*}

\subsection{Construction of Cardiac Meshes for CFD Simulations}

\subsubsection{Ablation Study of Individual Loss Components on Vessel Inlet/Outlet Structures}
\begin{table}[ht]
\caption{Ablation study of mesh regularization losses on vessel inlet and outlet structures on CT test dataset (N=20).}
\label{table:loss_ablation}
\resizebox{\columnwidth}{!}{%
\begin{tabular}{p{0.24\columnwidth}p{0.03\columnwidth}p{0.14\columnwidth}p{0.14\columnwidth}p{0.14\columnwidth}p{0.14\columnwidth}}
\toprule
                       &    & CoP+\par Ortho+HW & CoP+Ortho &         CoP &             None \\
\midrule
\multirow{3}{0.24\columnwidth}{Cap-Wall Orthogonality ($\downarrow$)} & LA &                   0.128$\pm$0.121 &     \textbf{0.032}$\pm$0.012 &  0.365$\pm$0.265 &  0.273$\pm$0.266 \\
                       & RA &                   0.023$\pm$0.008 &     \textbf{0.012}$\pm$0.008 &  0.105$\pm$0.038 &  0.066$\pm$0.026 \\
                       & Ao &                   0.019$\pm$0.023 &     \textbf{0.005}$\pm$0.006 &  0.467$\pm$0.117 &  0.127$\pm$0.024 \\
\cline{1-6}
\multirow{3}{0.24\columnwidth}{Cap Coplanarity (mm) ($\downarrow$)} & LA &                   0.228$\pm$0.041 &     0.256$\pm$0.029 &   \textbf{0.12}$\pm$0.024 &  0.312$\pm$0.058 \\
                       & RA &                    0.34$\pm$0.073 &     0.339$\pm$0.055 &  \textbf{0.185}$\pm$0.043 &  0.466$\pm$0.114 \\
                       & Ao &                   0.447$\pm$0.115 &     0.429$\pm$0.063 &  \textbf{0.263}$\pm$0.068 &   0.852$\pm$0.16 \\
\cline{1-6}
\multirow{2}{0.24\columnwidth}{Wall Chamfer Distance (mm) ($\downarrow$)} & LA &                   2.093$\pm$0.803 &     2.715$\pm$1.105 &  2.487$\pm$0.898 &  \textbf{2.042}$\pm$0.857 \\
                 & RA &                   2.021$\pm$1.176 &      2.66$\pm$1.301 &  2.231$\pm$0.983 &  \textbf{1.899}$\pm$0.952 \\
\bottomrule
\end{tabular}
}
\end{table}

\begin{figure}[ht]
    \centering
    \includegraphics[width=\columnwidth]{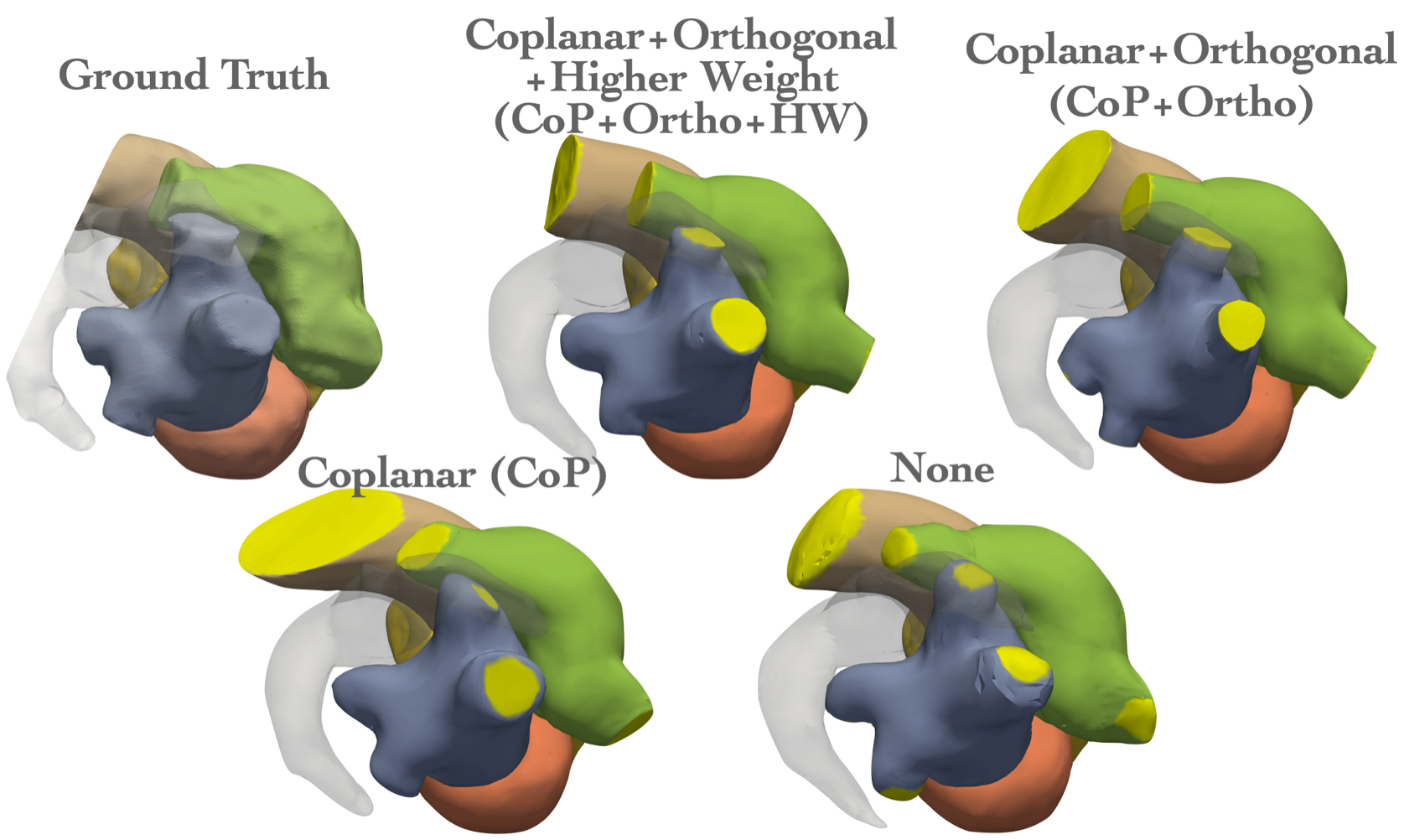}
    \caption{Visualization of example whole heart surface predictions following addition of regularization losses on vessel inlet/outlet structures. The yellow regions highlight the "caps" where the regularization losses were applied. }
    \label{fig:loss_ablation}
\end{figure}

CFD simulations of cardiac flow requires well-defined inlet and outlet vessel structures to prescribe boundary conditions for the inflow and outflow. Figure \ref{fig:loss_ablation} and table \ref{table:loss_ablation} demonstrate the effect of applying individual regularization loss components on the predicted inlet and outlet geometries (pulmonary veins, vena cava, and aorta). Without any of the regularization losses, the predicted vessel structure lacked well define caps. Indeed, our ground truth meshes were generated from manual segmentations where vessels were not truncated precisely orthogonal to the vessel walls by the human observers, and the caps were not co-planar due to necessary smoothing steps to filter out the staircase artifacts. The coplanar loss and the orthogonal loss succeeded in producing more planar cap geometries that were more orthogonal to vessel walls. Owning to the imperfect ground truth vessel meshes, although adding regularization losses to the training objective improved the structural quality of inlet geometries, it slightly reduced the geometric accuracy in terms of Chamfer distances compared with the ground truth. Applying a higher weight on the inlet mesh vertices in the geometric consistency loss was able to improve the geometric accuracy of vessel inlets while maintaining satisfactory mesh quality for CFD simulations.

\subsubsection{Comparison with Other Methods on Time-Series CT Data}
\begin{table}[]
\caption{Comparison of DCS ($\uparrow$) and HDs (mm) ($\downarrow$) of predictions from different methods on 4D CT test images (n=20). * Denotes Significant Difference Of "Ours (S)" From the Others (p<0.05)}\label{table:4dct}
\resizebox{\columnwidth}{!}{%
\begingroup
\setlength{\tabcolsep}{2pt}
\begin{tabular}{p{0.06\columnwidth}p{0.22\columnwidth}p{0.09\columnwidth}p{0.09\columnwidth}p{0.09\columnwidth}p{0.09\columnwidth}p{0.09\columnwidth}p{0.09\columnwidth}p{0.09\columnwidth}p{0.08\columnwidth}}
\toprule
        &               &              Myo &               LA &               LV &              RA &               RV &               Ao &              PA &               WH \\
\midrule
\multirow{4}{*}{Dice} & Ours (S) &            89.53 &            93.30 &            94.48 &           92.91 &            94.32 &            96.20 &  \textbf{85.31} &            93.14 \\
        & Ours (V) &           88.27* &           91.60* &           93.21* &          92.18* &           93.21* &           95.62* &          83.48* &           91.97* \\
        & HeartFFDNet &           84.37* &           88.38* &           91.41* &          90.26* &           90.19* &           93.03* &          70.44* &           88.94* \\
        & MeshDeformNet &  \textbf{90.58}* &  \textbf{95.18}* &  \textbf{95.85}* &  \textbf{93.50} &  \textbf{94.63}* &  \textbf{97.50}* &          80.21* &  \textbf{93.94}* \\
\cline{1-10}
\multirow{4}{*}{HD (mm)} & Ours (S) &             6.04 &            10.21 &             4.95 &  \textbf{10.04} &             6.61 &             3.80 &           19.71 &            16.02 \\
        & Ours (V) &   \textbf{5.91}* &            10.64 &             5.36 &           10.32 &            7.03* &             4.16 &  \textbf{19.14} &   \textbf{15.69} \\
        & HeartFFDNet &            6.78* &           12.01* &             6.37 &          10.85* &            8.77* &            5.13* &           23.20 &            16.46 \\
        & MeshDeformNet &             5.98 &    \textbf{9.29} &    \textbf{4.39} &           10.42 &   \textbf{6.35}* &    \textbf{3.42} &           23.25 &            15.77 \\
\bottomrule
\end{tabular}
\endgroup
}
\end{table}

\begin{figure}[ht]
    \centering
    \includegraphics[width=\columnwidth]{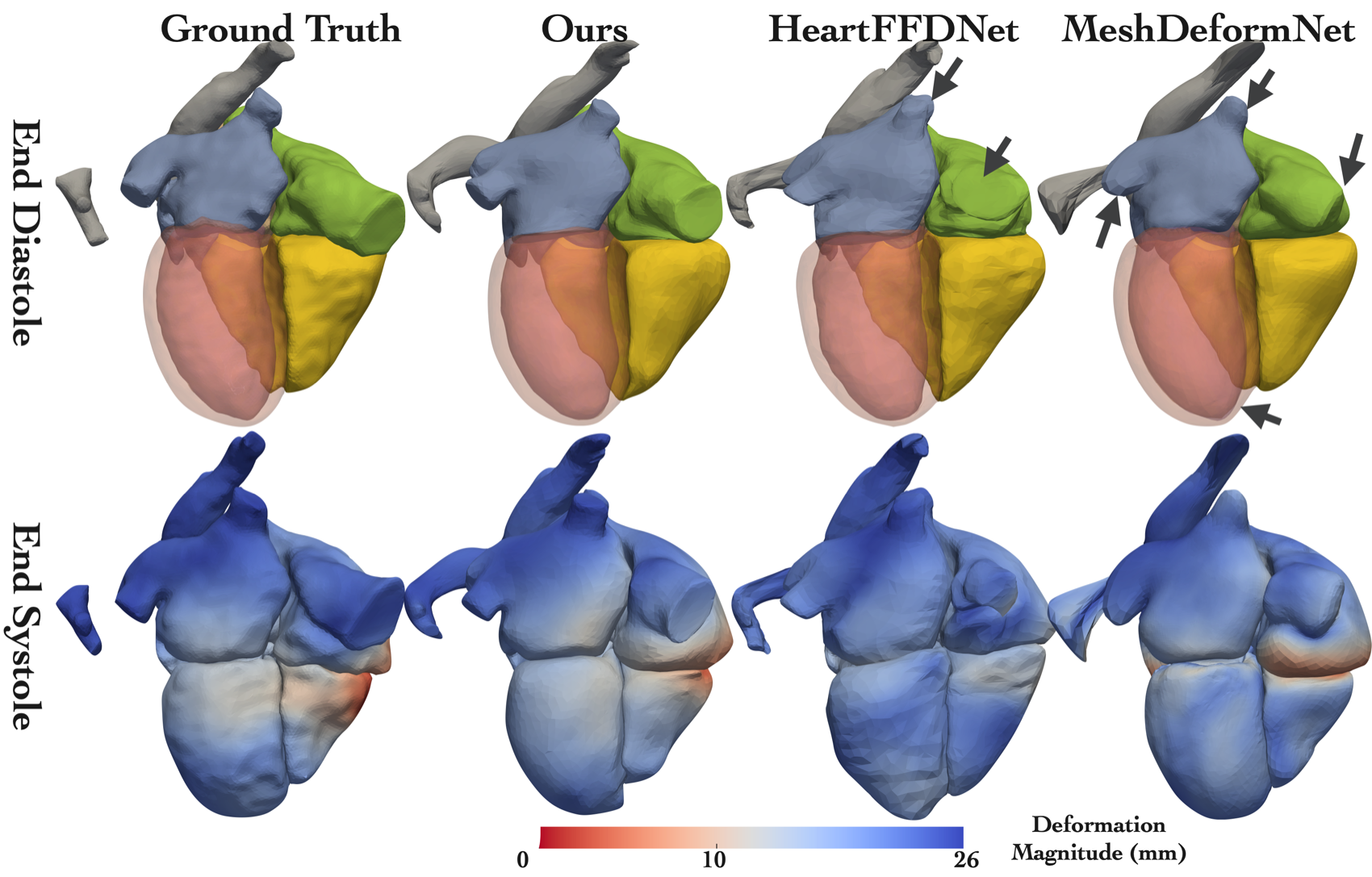}
    \caption{Qualitative comparison of whole heart surfaces from different methods at the end-diastolic phase and the end-systolic phase of a set of time-series image data. The colormap for end-systolic surfaces shows vertex displacement magnitude from end-systole to end-diastole.}
    \label{fig:example_mesh}
\end{figure}

Table \ref{table:4dct} compares the reconstruction accuracy between our method and the other baseline methods on end-diastolic and end-systolic phases of a cardiac cycle. Overall, our method demonstrated high accuracy comparable to the prior state-of-the-art approach MeshDeformNet, both in terms of Dice scores and Hausdorff distances. Figure \ref{fig:example_mesh} shows a qualitative comparison of the reconstructed whole heart surfaces at end-systolic and end-diastolic phases and the estimated surface motion by computing the displacements of mesh vertices over time. MeshDeformNet produced gaps between cardiac structures as well as overly smoothed pulmonary veins and vena cava geometries, since that method is biased by the use of sphere templates rather than a more fitting template of the whole heart. In contrast, our method produced high-quality geometries of the vessel inlets and outlets as well as whole heart geometries that better match with the ground truth. Furthermore, our method demonstrated a more accurate estimation of surface deformation over time, which is required for prescribing boundary conditions for CFD simulations. 

\begin{figure}[ht]
    \centering
    \includegraphics[width=\columnwidth]{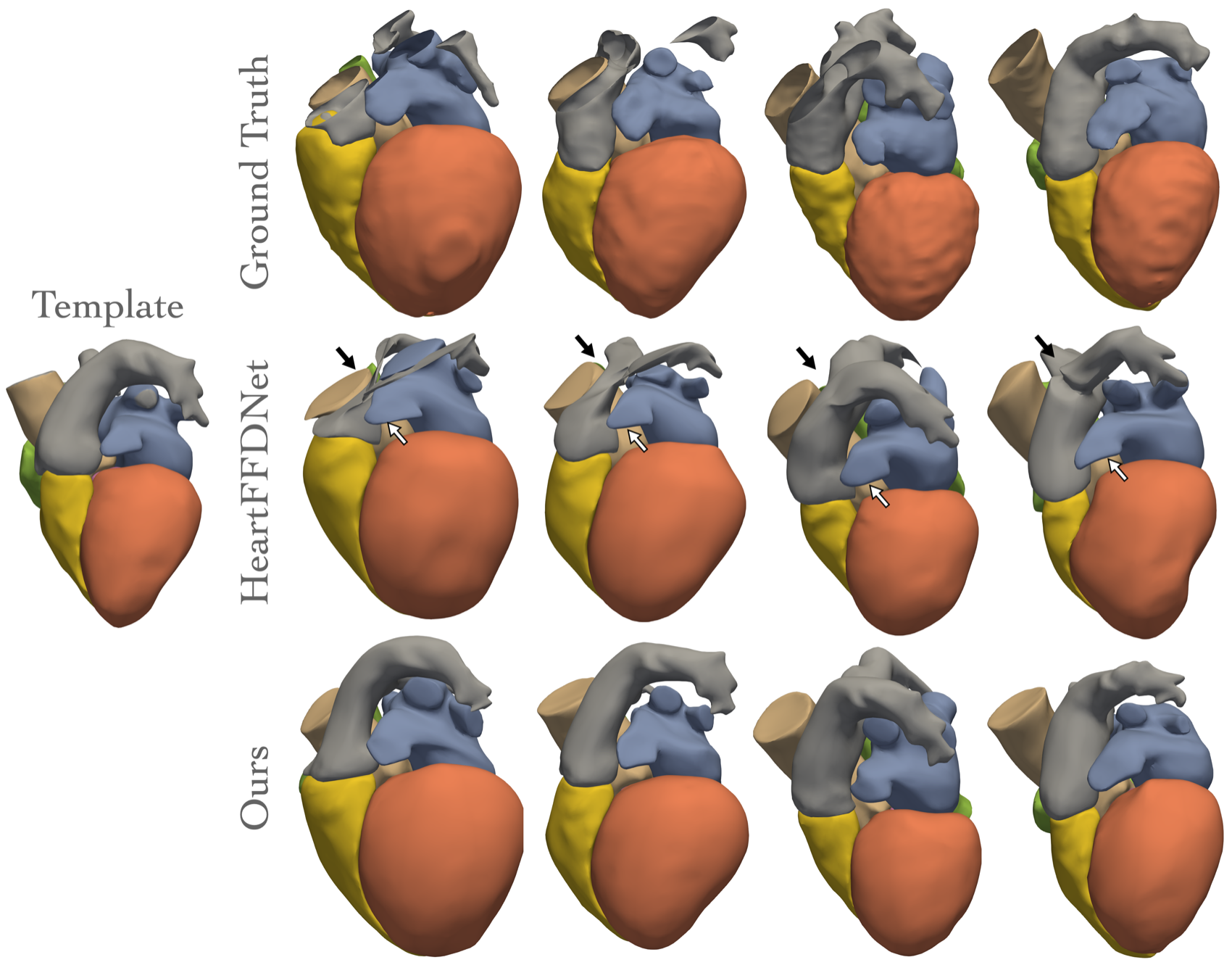}
    \caption{Comparison of whole heart surface predictions between using control handles as in our approach and using FFD as in HeartFFDNet.}
    \label{fig:bc_ffd}
\end{figure}

Figure \ref{fig:bc_ffd} provides further qualitative comparisons between using FFD and using biharmonic coordinates to deform the template. Using biharmonic coordinates enables more flexible deformation and can thus more closely capture detailed geometries such as the left atrial appendage. In contrast, geometries of left atrial appendage predicted from HeartFFDNet were strongly biased by the geometries of the template, although it used far more control points (4096) than our method (600). Furthermore, our method was able to predict cardiac structures that were not covered in the image data. Namely, thanks to the augmentation pipeline, our method generated reasonable geometries of the pulmonary arteries and pulmonary veins. In contrast, manual segmentation can only produce surface meshes of the cardiac structures captured in the images and HeartFFDNet predicted flat and unphysiological geometries despite starting from a realistic whole heart template. 

\begin{figure}[ht]
    \centering
    \includegraphics[width=\columnwidth]{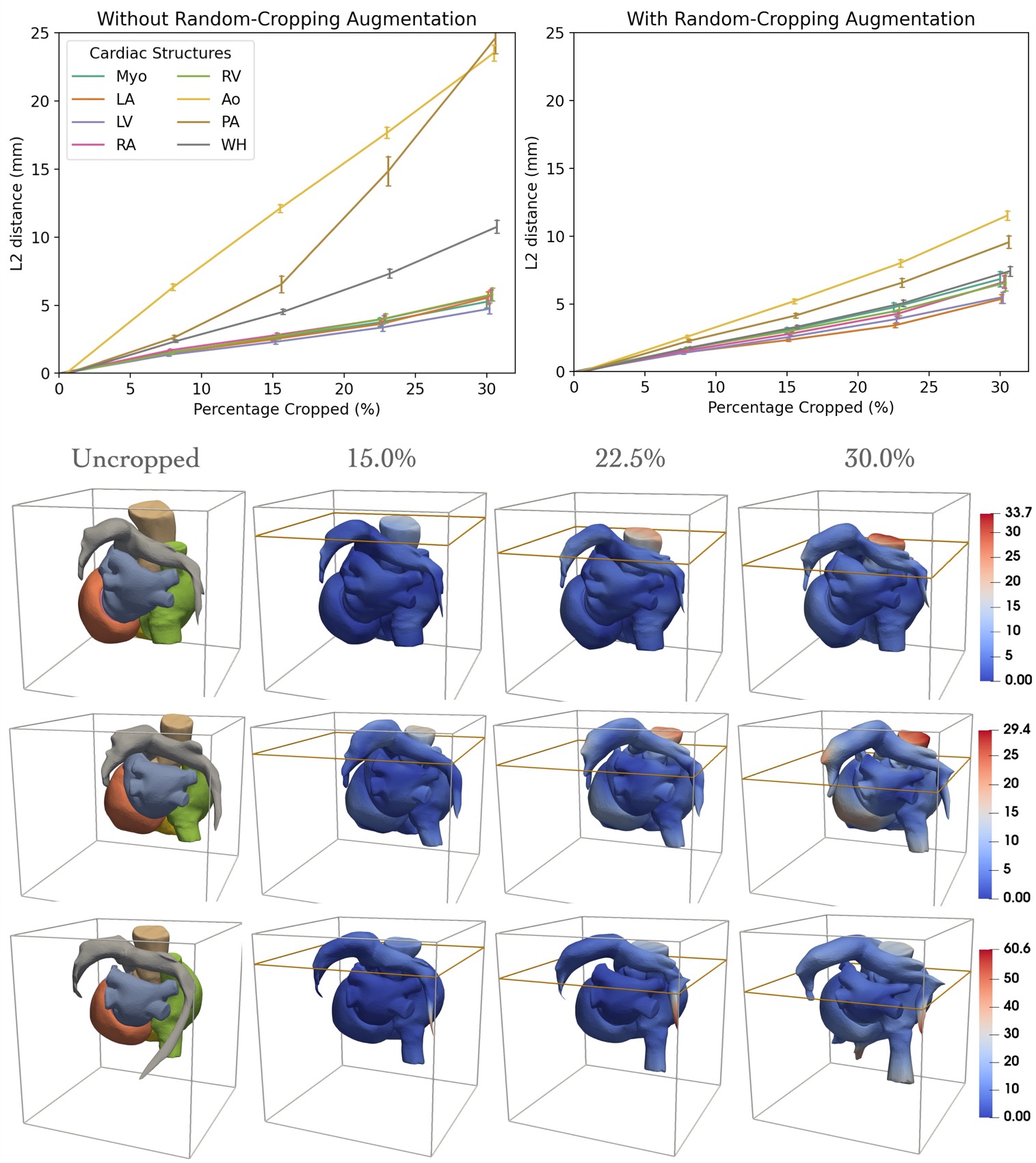}
    \caption{Accuracy of shape completion of the heart when image data has limited coverage for aorta, pulmonary arteries and veins. Top: A comparison of point-to-point L2 distance between the meshes predicted using uncropped input images and using cropped input images at varies percentage, for network models trained without and with the random-cropping augmentation. Bottom: Example prediction results for three cases using uncropped input images and using cropped input images at varies percentage. The color map on the meshes for uncropped images indicates different cardiac structures whereas those for cropped images indicates L2 distance in mm.}
    \label{fig:bc_completion}
\end{figure}

We further quantified the accuracy of the predicted shape of the heart outside the input image data. Since most images from the time-series CT dataset did not cover the entire heart, we selected 28 images that covered the entire heart from the MMWHS CT test dataset, and cropped them above various axial planes to evaluate the accuracy of the predicted whole heart reconstruction when increasing portions of cardiac structures were uncovered in the input images. Figure \ref{fig:bc_completion} (top) compares the average point-to-point distance errors for each cardiac structures when the image volume was cropped along the axial view before and after applying the proposed random-cropping augmentation method and weighted losses on the meshes. When the random-cropping augmentation was applied, our method produced more accurate aorta and pulmonary arteries. As expected, the distance errors increased when more image data were removed. When as much as 30\% of the image volume was removed, the average distance errors of pulmonary arteries and aorta were around 1 cm with the augmentation, whereas the average distance errors were around 2.5 cm without the augmentation. Figure \ref{fig:bc_completion} (bottom) visualizes three examples of the reconstruction results where $15\%$, $22.5\%$, and $30\%$ of the image data were removed. Our method produced reasonable geometries of the pulmonary arteries, pulmonary veins, and aorta in all cases. Although our method tended to predict a shorter aorta when the image data had limited coverage of the aorta, we note that the length of aortic outlet is often arbitrary when creating meshes for CFD simulations and thus does not significantly impact the simulation results. 

\begin{table*}[]
\caption{A comparison of the quality of the inlet/outlet geometries and whole heart surface quality from different methods.}\label{table:cap}
\resizebox{\textwidth}{!}{%
\begin{tabular}{llllllll}
\toprule
{} & \multicolumn{3}{c}{Cap-Wall Orthogonality ($\downarrow$)} & \multicolumn{3}{c}{Cap Coplanarity (mm) ($\downarrow$)} & \% Face Intersection ($\downarrow$) \\
{} &                     LA &               RA &               Ao &                     LA &               RA &               Ao &                   WH \\
\midrule
Ours (V)          &        \textbf{0.038}$\pm$0.046 &  \textbf{0.013}$\pm$0.007 &  \textbf{0.013}$\pm$0.012 &         \textbf{0.22}$\pm$0.024 &  \textbf{0.284}$\pm$0.044 &  \textbf{0.292}$\pm$0.088 &      \textbf{0.018}$\pm$0.022 \\
HeartFFDNet   &         0.137$\pm$0.08 &  0.228$\pm$0.182 &  0.494$\pm$0.386 &        0.398$\pm$0.068 &   0.45$\pm$0.125 &  0.949$\pm$0.557 &      0.262$\pm$0.191 \\
MeshDeformNet &        0.106$\pm$0.104 &  0.044$\pm$0.038 &  0.209$\pm$0.117 &        1.145$\pm$0.165 &  0.917$\pm$0.379 &   0.36$\pm$0.216 &      0.034$\pm$0.068 \\
\midrule
Manual        &          0.04$\pm$0.04 &  0.034$\pm$0.054 &  0.025$\pm$0.023 &        0.037$\pm$0.009 &  0.035$\pm$0.007 &   0.02$\pm$0.003 &          0.0$\pm$0.0 \\
\bottomrule
\end{tabular}
}
\end{table*}

Table \ref{table:cap} compares the quality of the predicted inlet and outlet geometries as well as the percentage face intersection of the whole heart meshes. Besides comparing with our baselines, HeartFFDNet and MeshDeformNet, we also compared our method with the surface meshes generated from applying the Marching Cube algorithm on manual ground truth segmentations, where the vessel inlet and outlet geometries were manually trimmed by human experts. Our method produced significantly better vessel inlet and outlet geometries than HeartFFDNet and MeshDeformNet. Also, our method outperformed the manual segmentation in terms of Cap-Wall Orthogonality. When deforming the mesh template using control handles, our method achieved the lowest percentage of face intersection than other deep-learning methods, and the small amount of face intersections that occurred could be readily corrected by a few iterations of Laplacian smoothing. 

\subsection{CFD Simulations of Cardiac Flow}

\begin{figure}[ht]
    \centering
    \includegraphics[width=\columnwidth]{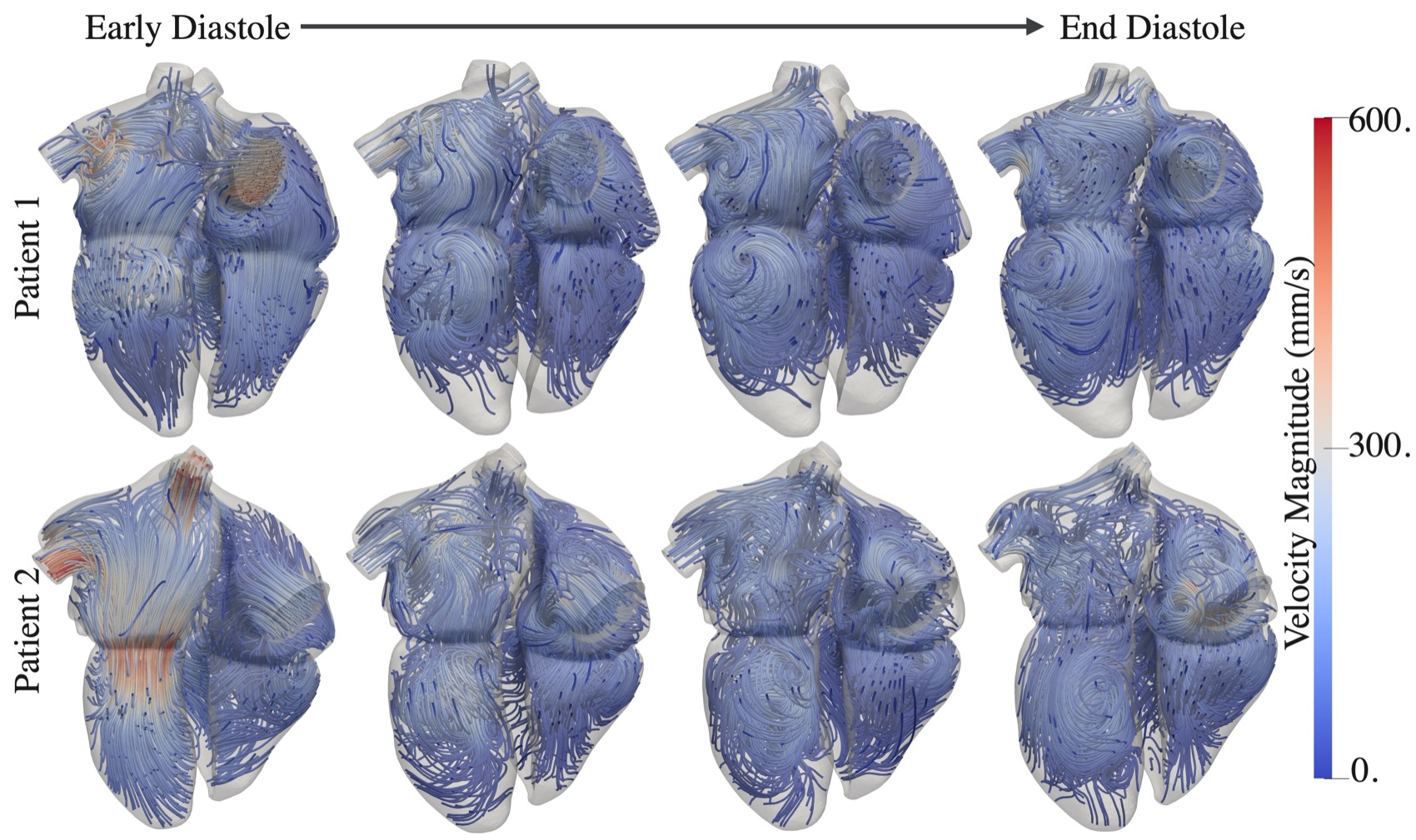}
    \caption{Velocity streamlines from CFD simulations of 2 different patients using the predicted 4D meshes.}
    \label{fig:cfd_wh}
\end{figure}

\textcolor{black}{We were able to successfully conduct CFD simulations using the automatically constructed LV meshes for all 10 patients, as well as for 9 of 10 patients with the 4-chamber meshes. The 1 failed case had structure penetrations between two pulmonary veins, causing the simulation to diverge. Figure \ref{fig:cfd_wh} displays the simulation results of the velocity streamlines at multiple time steps during diastole for 2 different patients. The simulation results demonstrate the formation of typical vortex flow during ventricle filling.} 

\begin{figure}[ht]
    \centering
    \includegraphics[width=\columnwidth]{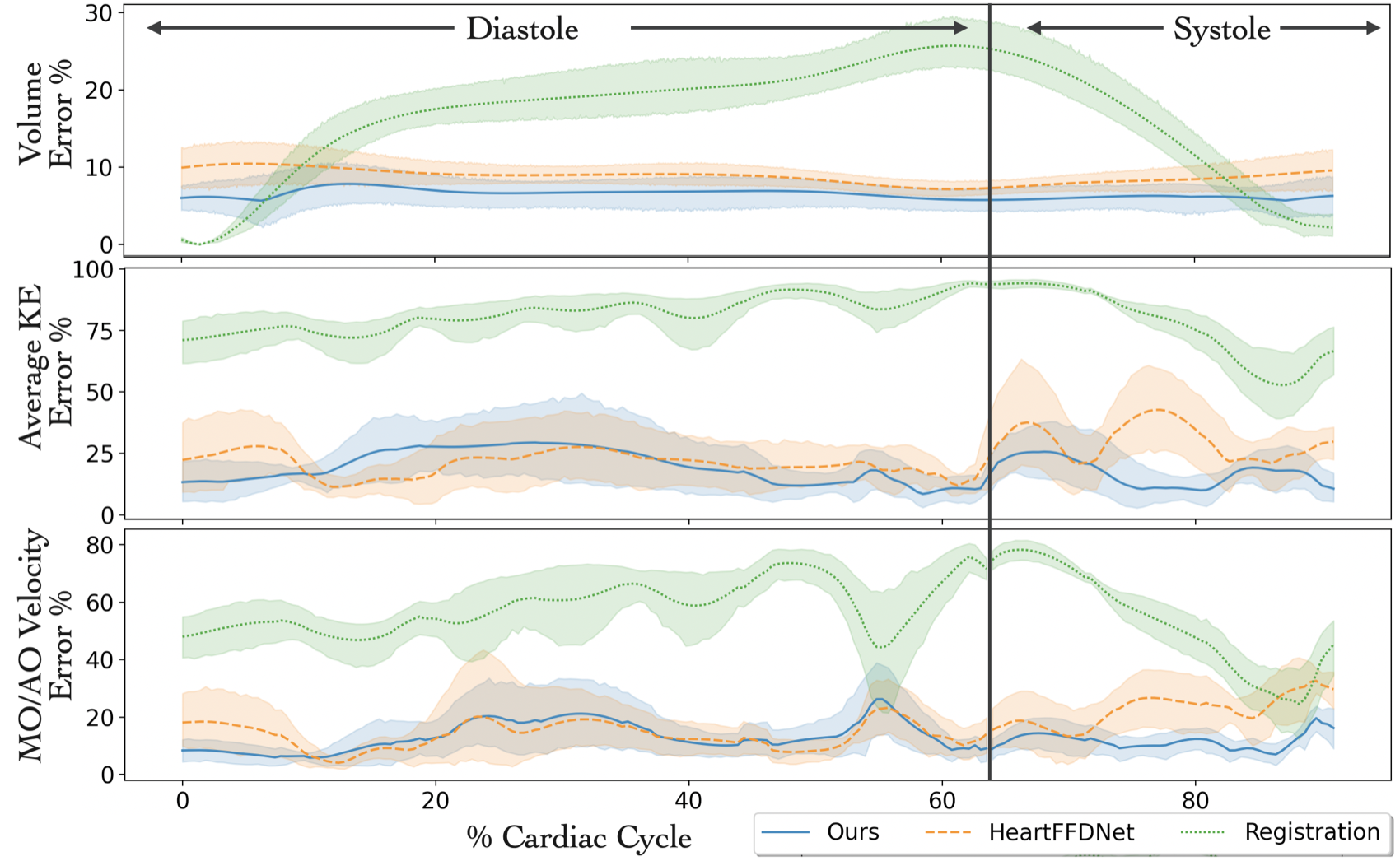}
    \caption{Quantitative comparisons of the \% errors in LV volume, volume averaged KE density, mean velocity near the MO during diastole and mean velocity near the AO during systole among different methods. Lines show the mean values and shades show the 95\% confidence intervals.}
    \label{fig:cfd_quantitative}
\end{figure}

Figure \ref{fig:cfd_quantitative} provides quantitative comparisons of the accuracy of CFD simulation results of LV flow. Both our approach and HeartFFDNet significantly outperformed the image-registration-based approach in terms of all metrics. Namely, the image-registration-based method significantly underestimated the LV volume during diastole since the reconstructed meshes did not capture the large deformation of LV from systole to diastole. Our proposed approach demonstrated comparable or slightly better accuracy than HeartFFDNet in general, with smaller volume errors throughout the cardiac cycle and smaller errors in average kinetic energy and mean aortic flow velocity during systole. 

\begin{figure*}[ht]
    \centering
    \includegraphics[width=\textwidth]{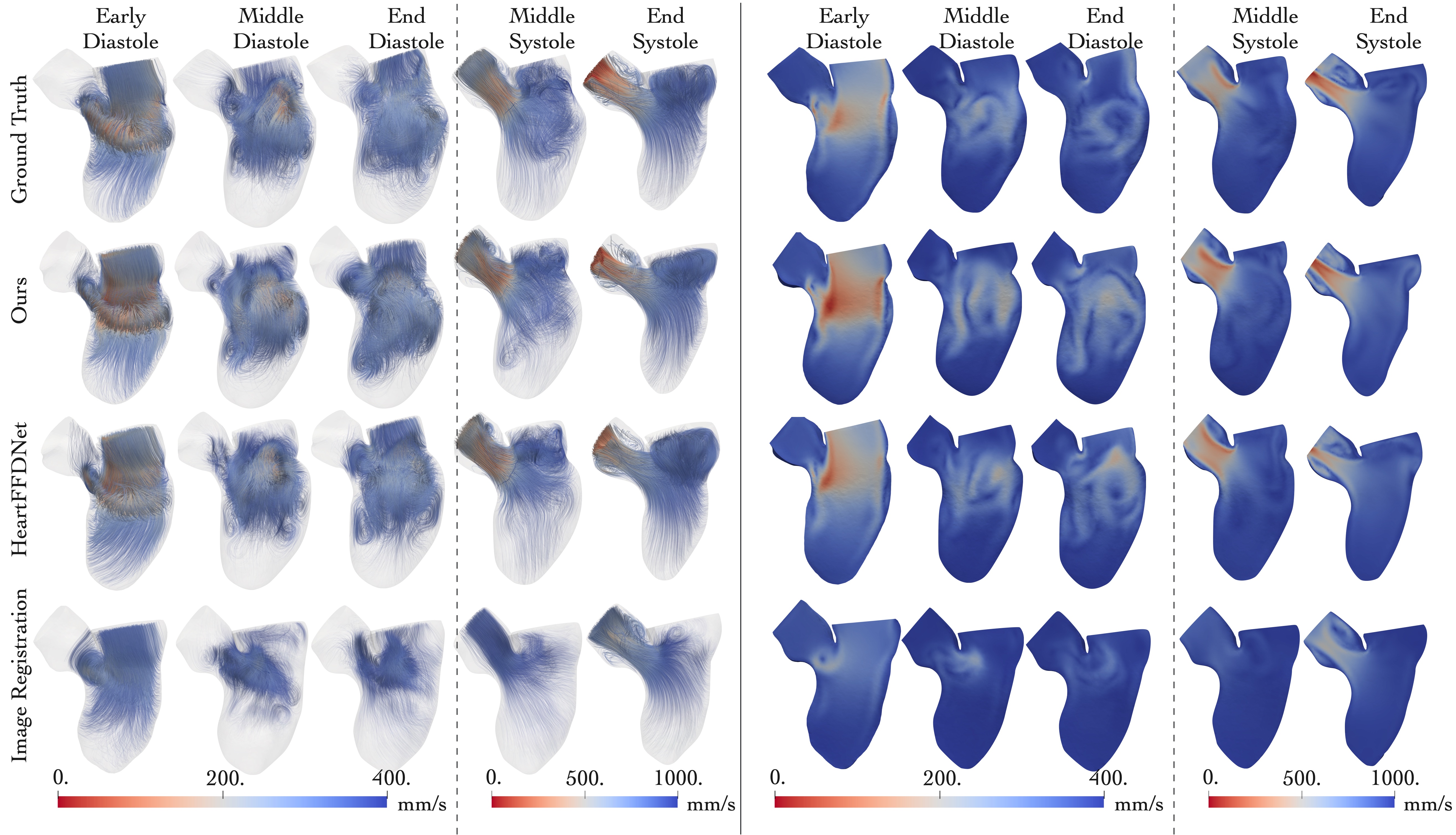}
    \caption{Qualitative comparisons of the simulated flow pattern from different methods at different time phases during a cardiac cycle for an example case. Left: Streamlines within the left ventricle models. Right: Contours of the velocity magnitude at the same clipping plane. Color map shows the velocity magnitude}
    \label{fig:cfd_compare}
\end{figure*}

Figure \ref{fig:cfd_compare} qualitatively compares the simulated LV flow pattern during both systole and diastole using meshes automatically constructed by our proposed approach and HeartFFDNet, semi-automatically constructed by conventional image registration and manually constructed by human observers. Image registration underestimated the LV expansion from end systole to diastole, leading to underestimated flow velocity and disparate flow pattern compared with the ground truth. Both of our approaches generally produced similar vortex structures during diastole and converging flow during systole, with moderate differences in flow velocity and vortex locations compared with the ground truth.

\section{Discussion}

Automated image-based reconstruction of cardiac meshes is important for computational simulation of cardiac physiology. While deep-learning-based methods have demonstrated success in tasks such as image segmentation and registration, few studies have addressed the end-to-end learning between images and meshes for modeling applications. Furthermore, prior learning-based mesh reconstruction approaches suffer from a number of limitations such as using decoupled meshes of individual cardiac structures and assumed mesh topology, thus unable to directly support different cardiac simulations without additional efforts \cite{Voxel2Mesh, kong2021deeplearning}. We addressed this challenge herein using a novel approach that trains a neural network to learn the translation of a small set of control handles to deform the space enclosing a whole heart template to fit the cardiac structures in volumetric patient image data. Our method demonstrated promising whole-heart reconstruction accuracy and was able to generate simulation-ready meshes from time-series image data for CFD simulations of cardiac flow.

Our approach achieved comparable geometric accuracy to the prior state-of-the-art whole heart mesh reconstruction method MeshDeformNet \cite{kong2021deeplearning} while having the additional advantage of directly enabling various cardiac simulations. We note that our approach used fewer parameters in the CNN encoder compared to MeshDeformNet (Table \ref{table:size}) and the use of biharmonic coordinates naturally ensures the smoothness of deformation without using explicit mesh regularization (e.g., Laplacian and/or edge length loss constraints \cite{kong2021deeplearning}). This is important since mesh regularization schemes can complicate the optimization process \cite{NeuralMeshFlow}, whereas we observed our approach to converge significantly faster than MeshDeformNet (18 vs 32 hrs on a GTX2080Ti GPU). 

\begin{table}[]
\caption{Comparison of model size, training and testing time.}
\label{table:size}
\resizebox{\columnwidth}{!}{
\begin{tabular}{lccccc}
\hline
                    & Ours  & HeartFFDNet & MeshDeformNet & 2D UNet & 3D UNet \\ \hline
\# of Parameters & 8.7M   & 8.5M         & 16.8M         & 31.1M   & 18.6M    \\
Training Time & 18 hrs    & 26 hrs          & 32 hrs            & 7 hrs      &  37 hrs       \\
Test Time     & 0.230s & 0.177s       & 0.425s        & 1.555s  & 0.367s  \\ \hline
\end{tabular}
}
\end{table}

For CFD simulations requiring the time-dependent motion of the heart over the cardiac cycle, our method has the advantage of deforming the template mesh in a temporally consistent manner, enabling automated construction of dynamic cardiac meshes within minutes on a standard desktop computer. Registration-based approaches, in contrast, often require test time optimizations that are computationally expensive and prone to local minimums, which often lead to inaccurate registration results such as underestimation of large deformation. Although deep-learning approaches have been proposed to speed-up the registration process \cite{Balakrishnan2019VoxelMorphAL, Mok2020LargeDD}, large-deformation registration on cardiac images remains challenging for learning-based approaches. The establishment of temporal feature correspondence of our method is due to similar features of time-series images naturally being encoded into similar feature vectors by the CNN encoders and does not require explicit training. Nevertheless, ground truth data of anatomical landmarks could be incorporated in the future during training to further improve the accuracy of feature correspondence across different time frames or patients.

Blood flow simulations developed from our automated mesh generation process demonstrated circulatory flow patterns during diastole and converging flow patterns during systole in the ventricular cavity consistent with prior studies \cite{Vedula2015, Mittal}. However, we observed an average of 15-25\% error in the simulated mean velocity and kinetic energy, despite a promising mean LV Dice scores of 93\% and an mean volume error of 6\%. Such amount of volume error is consistent with the inter- and intra-observer variations of manual LV segmentation\cite{ZHUANG2019, kong2021deeplearning}. Indeed, simulation of blood flow is sensitive to uncertainties in geometry, such as inflow directions and vessel and/or LV wall smoothness \cite{Vedula2015,Celi2021OnTR}. We plan to conduct intra- and inter-observer studies on the ground truth meshes to further understand the relationship between prediction uncertainties and the accuracy of CFD simulations. Nevertheless, our approach is among the first to enable creation of simulation-suitable meshes from patient images. And our design of using template meshes and control handles could support shape editing and analysis to study the effect of geometric variations on CFD simulations.

Our proposed method has the following limitations. First, the  testing images we used from the benchmark MMWHS test dataset and the times-series CT dataset do not cover the full variations of cardiac abnormalities observed clinically. For example, our test datasets did not contain patients with congenital heart defects and thus the performance of our trained network for those patients were not evaluated. Furthermore, image data used to training and testing  using relatively similar imaging protocols and parameters, such as slice thickness, field of view, and spatial orientation. The above limitations can be addressed by retraining the model with data more representative of particular use cases (e.g. particular abnormalities or scanner protocol). A related limitation is that the whole-heart template used may need to be modified to capture richer variations of cardiac malformations. For example, the current template assumes four separate and distinct pulmonary vein ostia and thus may not fully capture pulmonary veins with alternate branching patterns, which can be important for preoperative planning of pulmonary and cardiac surgery \cite{Kandathil2018PulmonaryVA}. Similarly, the template used would not be suitable for cardiac malformations such as single-ventricle patients with congenital heart diseases since the structures of the heart are significantly different from our current training template. Nonetheless, this framework could still be utilized if sufficient training data of, say, single-ventricle patients were available and a corresponding single-ventricle mesh template were used. To better handle the above applications, in future work we aim to add a template retrieval module to automatically select a template that best suits the application. Furthermore, implicit shape representation \cite{Deng2021DeformedIF} can be combined with our learning-based shape deformation approach to predict cardiac structures with different anatomies.

\section{Conclusion}
We proposed a novel deep-learning approach that directly constructs simulation-ready whole heart meshes from cardiac image data and allows switching of template meshes to accommodate different modeling requirements. Our method leverages a graph convolutional network to predict the translations of a small set of control handles to smoothly deform a whole heart template using biharmonic coordinates. Our method consistently outperformed prior state-of-the-art methods in constructing simulation-ready meshes of the heart, and was able to produce geometries that better satisfy modeling requirements for cardiac flow simulations. We demonstrated application of our method on constructing dynamic whole heart meshes from time-series CT image data to simulate the ventricular flow driven by the cardiac motion.  The presented approach is able to automatically construct whole heart meshes within seconds on a modern desktop computer and has the potential in facilitating high-throughput, large-cohort validation of patient-specific cardiac modeling, as well as its future clinical applications.

\bibliographystyle{IEEEtran}
\bibliography{ref}
\end{document}